\documentclass[12pt]{article}
\usepackage[utf8]{inputenc}
\usepackage{empheq}
\usepackage{bm}
\usepackage[normalem]{ulem}
\usepackage{latexsym}
\usepackage{color,multirow}
\usepackage{amsmath,amsfonts,amssymb,geometry}
\usepackage{graphicx,epsfig, cite,tocloft}
\usepackage{psfrag}
\usepackage{amsthm}
\usepackage{cite,url}
 \usepackage{makecell}
 \usepackage{hyperref}
\pdfoutput=1
\usepackage{amsmath}
\usepackage{amssymb}
\usepackage{amsfonts}
\usepackage{mathrsfs}
\usepackage{mathrsfs}
\usepackage{fullpage}
\usepackage{setspace}
\usepackage{bm}
\usepackage{bbm}
\usepackage[normalem]{ulem}
\usepackage{enumerate}
\usepackage{yfonts}
\usepackage{psfrag}

\usepackage{graphicx}
\usepackage{cancel}
\usepackage{slashed}

\usepackage[font=small,labelfont=bf]{caption}
\usepackage{subcaption}

\newcommand{\be}{\begin{equation}}
\newcommand{\bea}{\begin{eqnarray}}
\newcommand{\ee}{\end{equation}}
\newcommand{\eea}{\end{eqnarray}}

\begin{document}
\numberwithin{equation}{section}
{
\begin{titlepage}
\begin{center}
\hfill \\
\hfill \\
\vskip 0.2in
 {\Large{Analytical study of gravitational lensing in Kerr-Newman black-bounce spacetime}}\\
\vskip .7in

{\large Saptaswa Ghosh${}$\footnote{\url{saptaswaghosh@iitgn.ac.in}},
 Arpan Bhattacharyya${}$\footnote{\url{abhattacharyya@iitgn.ac.in}}}

\vskip 0.3in

{\it Indian Institute of Technology, Gandhinagar, Gujarat-382355, India}\vskip .5mm

\vskip.5mm

\end{center}

\vskip 0.35in

\begin{center} {\bf ABSTRACT } \end{center}
We investigate the equatorial deflection angle of light rays propagating in Kerr-Newman black-bounce spacetime. Furthermore, we analyze the light ray trajectories and derive a closed-form formula for deflection angle in terms of elliptic integrals. The deflection angle increases with the decrease of charge and regularisation parameter for a particular impact parameter. We also study the strong field limit of the deflection angle. Using this strong deflection angle formula and lens equation, we find the radius of the first Einstein ring and study its dependence on the charge and the regularisation parameter. We demonstrate that the charge has a robust effect on the size of the Einstein rings, but the effect of the regularization parameter on the ring size is negligible. We also investigate the non-equatorial lensing and the caustic structures for small polar inclination, and the same observations appear to hold. These results directly affect the observational appearance of the Kerr-Newman black-bounce.
\vfill
%\noindent \today
\end{titlepage}
}
%%%%%%%%%%%%%%%%%%%%%%%%%%%%%%%%%%%%%%%%%%%%%%%%%%%%%%%%%%%%%%%%%%%%%%%%%%%%%%%%
% Table of contents
%%%%%%%%%%%%%%%%%%%%%%%%%%%%%%%%%%%%%%%%%%%%%%%%%%%%%%%%%%%%%%%%%%%%%%%%%%%%%%%%
\newpage
\tableofcontents
%%%%%%%%%%%%%%%%%%%%%%%%%%%%%%%%%%%%%%%%%%%%%%%%%

\section{Introduction}

Black holes arising out as solutions to Einstein's theory of General Relativity are an important testbed to probe into the intricate structures of spacetime. However, finding such objects was a great challenge until EHT gave us the first glimpse. Going on to the other way of getting data regarding such objects also gives us insight into which gravity theory observation of such objects provides essential data to ensure we are on the correct path. However, the recent developments in gravitational-wave astronomy ensure that such entities exist in nature. The shreds of evidence come from the set of detections of gravitational waves from binary black holes by LIGO, Virgo and KAGRA Collaboration\cite{LIGOScientific:2016aoc,LIGOScientific:2016vlm, LIGOScientific:2016sjg, LIGOScientific:2017bnn,LIGOScientific:2021qlt,Kokeyama:2020dkg}. Also, we have enough observational evidence that ensures the existence of supermassive black holes at the centre of our Milky Way galaxy. In recent times, ``Event Horizon Telescope" (EHT) collaboration has given us the observational signatures of such kinds supermassive black holes \cite{EventHorizonTelescope:2019dse,EventHorizonTelescope:2019uob,EventHorizonTelescope:2019jan, EventHorizonTelescope:2019ths,EventHorizonTelescope:2019pgp, EventHorizonTelescope:2019ggy, EventHorizonTelescope:2022xnr, EventHorizonTelescope:2022vjs, EventHorizonTelescope:2022wok, EventHorizonTelescope:2022exc, EventHorizonTelescope:2022urf, EventHorizonTelescope:2022xqj,
EventHorizonTelescope:2021dqv}. Together with the data from LIGO-Virgo-KAGRA, these observations provide a unique opportunity to test various aspects of our theoretical predictions, particularly in the strong gravity regime \cite{LIGOScientific:2019fpa,LIGOScientific:2021sio,Psaltis:2018xkc,Himwich:2020msm, Gralla:2020pra,Vagnozzi:2022moj,Volkel:2020xlc,Bambi:2019tjh,Vagnozzi:2019apd,Allahyari:2019jqz,Khodadi:2020jij}. \par
The deflection of light in a gravitational field is one of the significant consequences of General Relativity, technically called gravitational lensing. One typically uses geometrical optics approximation to compute the deflection angle \cite{Virbhadra:1999nm}. It can be shown that the deflection angle of light by some astrophysical objects depends on the distance between the object and the observer. Also, it depends on the parameters of underlying astrophysical objects. So, by calculating the deflection angle, one can estimate the parameters (spin, charge, mass), and by matching these with the observational data, one can validate a theory of gravity. It could provide observational constraints on theories beyond general relativity in some regimes \cite{Bhadra:2003zs,PhysRevD.80.024036,Eiroa:2005ag}. The calculation of the exact deflection angle is complicated for an arbitrary metric. Hence, we must make assumptions by considering what kind of object we are looking for \cite{Bartelmann:2016dvf, Brainerd:1995da,Virbhadra:1999nm,Frittelli:1999yf,Bozza:2002zj}.\par
The theory of gravitational lensing in the weak field limit was already developed in \cite{Book} long ago. It helps us estimate the parameters of less massive objects like stars and galaxies. However, the treatment fails when the objects are very compact and massive. Then we need to consider the strong field limit of deflection angle. In recent days, investigating the strong field limit of bending angle has become crucial since it could provide us with fundamental properties of supermassive objects like black holes. Based on the work \cite{Virbhadra:1999nm} and  \cite{Frittelli:1999yf} an analytical description of the deflection angle in the strong gravitational field has been proposed in \cite{Bozza:2002zj}, where the authors have shown the deflection angle logarithmically diverges when the turning point of the light rays gets close to the photon sphere. In \cite{Bozza:2002zj}, this analysis was presented for spherically symmetric black hole. Furthermore, this work was extended for rotating black hole in \cite{Bozza:2002af}. Treating black holes as lens, a lens equation has been derived in \cite{Virbhadra:1999nm} and further corrected and generalized in \cite{Bozza:2008ev}. Further analysis, including investigation of  Einstein rings \cite{Einstein:1936llh}, of the lens equation, and strong gravitational lensing based on the work \cite{Bozza:2002af}, has been made for various black hole spacetimes and compact objects \cite{Eiroa:2010wm,Bin-Nun:2010lws,Amarilla:2010zq,Stefanov:2010xz, Wei:2011nj,Chen:2011ef,Gyulchev:2012ty,Tsukamoto:2012xs,Sahu:2012er,Chen:2012kn,Wei:2013kza,Atamurotov:2013sca,Eiroa:2013nra,Wei:2014dka,Tsukamoto:2014dta,Liu:2015zou,Wei:2015qca,Sotani:2015ewa,Bisnovatyi-Kogan:2015dxa,Sharif:2015qfa,Younas:2015sva,Chen:2015cpa,Schee:2017hof,Man:2012ivp,Tsukamoto:2016jzh,Wang:2016paq,Tsukamoto:2016qro,Zhao:2016ltm,Aldi:2016ntn,Lu:2016gsf,Cavalcanti:2016mbe,Zhao:2016kft,Shaikh:2017zfl,Rahman:2018fgy,Virbhadra:2022iiy,Hsieh:2021scb,Hsieh:2021rru,Edery:2006hm}. Another consequence of the strong deflection analysis is the investigation of black hole shadow, a two-dimensional dark zone in the celestial sphere caused by the strong lensing of the black hole, which is a dark silhouette of the BH against a bright background. Although we will use this particular notion of shadow in this paper following \cite{EventHorizonTelescope:2019dse}, there are other formulations to quantify the visual appearance of a black hole. Interested readers are referred to this review \cite{Perlick:2021aok} for more details. Together with the Einstein ring, the structure of the shadow can constrain the underlying parameters (e.g. charge, spin, mass) of the spacetime and, consequently, the parameters of the underlying theory \cite{Psaltis:2018xkc,Himwich:2020msm,Gralla:2020pra,Vagnozzi:2022moj,Volkel:2020xlc} well as their observational implications, have been investigated in recent times \cite{Cunha:2015yba,Wang:2017hjl,Younsi:2016azx,Bisnovatyi-Kogan:2018vxl, Banerjee:2019nnj,Tsupko:2019mfo,Tsupko:2019pzg, Mishra:2019trb, Vagnozzi:2020quf,Li:2020drn,Perlick:2017fio,Wei:2018xks,Chowdhuri:2020ipb,Papnoi:2014aaa,Amarilla:2013sj,Amarilla:2011fx,Perlick:2021aok,Sarkar:2021djs,Lin:2022ksb,Chandrasekhar:579245,Adler:2022qtb,Virbhadra:2022ybp,Bambi:2019tjh,Vagnozzi:2019apd,Allahyari:2019jqz,Khodadi:2020jij,Roy:2021uye,Chen:2022nbb}  \footnote{This list is by no means exhaustive. Interested readers are referred to this review \cite{Perlick:2021aok} and citations there for more details.}.
\par
Recently, a static black-bounce metric that can describe a black hole and a wormhole spacetime under suitable circumstances has been suggested in \cite{Simpson:2018tsi ,Lobo:2020ffi, Mazza:2021rgq, Franzin:2021vnj}. An interesting feature of this spacetime \cite{Simpson:2018tsi}, is that it doesn't have a central singularity since an extra parameter has regularised it. This metric is not a solution to the vacuum Einstein equation. It requires the presence non-trivial energy-momentum tensor. The black-bounce spacetime has two different kinds of solutions depending on the choice of underlying regularisation parameters. When the solution does not admit any horizon, it corresponds to a Traversable wormhole solution. Otherwise, one can have a regular black hole solution without singularity (sometimes called a hidden wormhole), but there are two copies of it. Later, using the Newman-Janis algorithm, a rotating counterpart of this static black hole solution has been proposed in \cite{Mazza:2021rgq}. \par 
 In recent times, the gravitational lensing in strong deflection limit and shadows for Schwarzschild and Reissner-Nordstorm black-bounce metrics separately have been investigated in \cite{Tsukamoto:2020bjm, Nascimento:2020ime, Guerrero:2021ues}. Also, the strong deflection analysis for rotating black-bounce spacetime has been analyzed in \cite{Islam:2021ful} and the shadow of charged black-bounce metric has been analyzed in \cite{Guo:2021wid}. The weak lensing in the context of charged black-bounce metric has been investigated in \cite{Zhang:2022nnj}. In this paper, we consider the most general scenario by studying a charged rotating black-bounce metric \cite{Franzin:2021vnj}. First and foremost, we present an analytical computation of deflection angle for this charged black-bounce spacetimes, which has not been done. Then we consider the strong deflection limit and discuss the dependence of the radius of the Einstein ring on the charge and the regularisation parameter. Lastly, we compute various observable related to the shadow cast by this metric and discuss their dependence of them on the underlying parameters of the spacetime. Particularly motivated by the recent observation from EHT \cite{EventHorizonTelescope:2021dqv}, using the size of the shadow one perhaps can constrain the charge of the underlying spacetime, we have investigated how the observables related to the shadow for this charged rotating black-bounce depends on the charge parameter. \par
In this paper, we investigated the general setup that is possible. The article is organized as follows. In Sec.~(\ref{sec2}), we review the null geodesics in a charged, rotating black-bounce spacetime(also known as Kerr-Newman black-bounce spacetime) and find out the turning point of light rays and how it depends on the impact parameter. We also investigate the critical impact parameter and the innermost circular orbit in this section. In the Sec.~(\ref{sec3}), we give a general perturbative method to calculate the equatorial deflection angle. Next, in Sec.~(\ref{sec4}), we briefly consider the strong deflection analysis of equatorial lensing. In Sec.~(\ref{sec5}), we discuss some observational signatures and their dependence on charge and regularisation parameters in a strong deflection limit. In Sec~(\ref{Sec6}) we discuss the strong lensing in the non-equatorial plane  at small inclination and comment on the  caustic structure. Finally, we discuss observable quantities related to shadow size, their dependence on charge and regularisation parameters, and the implications of our results in Sec.~(\ref{sec6}). In Sec.~(\ref{sec7}), we give concluding remarks. Some necessary mathematical details are given in the Appendix ~(\ref{App}). Also, we have set the value of the speed of light $c$ and Newton's gravitational constant $G$ to unity.

\section{Null geodesics in the Kerr-Newman black-bounce spacetime}\label{sec2}
 We briefly discuss the null geodesics in Kerr-Newman black-bounce spacetime. In Boyer-Lindquist coordinate the line element can be written as \cite{Franzin:2021vnj}, 
\begin{eqnarray}
    ds^2&=&g_{{\mu}{\nu}} dx^{\mu}dx^{\nu}\nonumber\\&=&-\frac{\Delta}{\rho^2}(a\, \sin^2 \theta\, d\phi-dt)^2 + \frac{\sin^2\theta}{\rho^2}((r^2+a^2+l^2)d\phi -a\, dt)^2+ \rho^2 (\frac{dr^2}{\Delta}+d\theta^2)\label{21}
\end{eqnarray}
where
\begin{equation}
 \rho^2=r^2+l^2+a^2 \cos^2\theta\\,
 \Delta(r)=(r^2+a^2+l^2)-2m\sqrt{r^2+l^2}+Q^2\,. \label{22}
 \end{equation}
 In (\ref{21}) and (\ref{22}), $m\ge0$ is the ADM mass, $Q$ is the charge of the black hole, $a=\frac{J}{m}$ is the angular momentum per unit mass and $l>0$ is the extra parameter that is the cause of the non-existence of the central singularity. Note that in this case, the range of the radial coordinate is $-\infty<r<\infty$. The Kerr-Newman black-bounce metric reduces to the Kerr metric by taking $Q=0,l=0$.
The event horizon can be found by equating $\Delta(r)=0$ and is given by,
\begin{eqnarray}
     R_H&=&\sqrt{\Big[\Big(m+\sqrt{m^2-(a^2+Q^2)}\Big)^2-l^2\Big]}
\end{eqnarray}
along with the reality condition $m^2-(a^2+Q^2)>0$ and $m+\sqrt{m^2-(a^2+Q^2)}>l$. The lagrangian for photon is given by,
\begin{eqnarray}
   \mathcal{L}=\frac{1}{2}g_{\mu\nu}\dot{x}^{\mu}\dot{x}^{\nu}
\end{eqnarray}
where the four-velocity, $\dot{x}^{\mu}=\frac{dx^\mu}{d\tilde\lambda}$, is defined in terms of some convenient parameter $\tilde \lambda$.\par
Light rays travels along null geodesic obeying the condition $\dot{x}^{\mu}\dot{x}_{\mu}=0$.
As the consequences of the fact that the Kerr-Newman black-bounce spacetime is independent of time $t$ and azimuthal coordinate $\phi$, there exists two associated killing vectors $\xi_t^{\mu} $=(1,0,0,0) and $ \xi_\phi^{\mu}$=(0,0,0,1). The associated conserved quantities are: $E=-\xi_t^\mu \dot{x}_\mu\, \text{and}\, L=\xi_\phi ^\mu \dot{x}_\mu$.
Additionally there exists a Carter constant $\mathcal{K}$ coming from the separability of Hamilton-Jacobi equation and is given by,
\begin{eqnarray}
   \mathcal{K}&=&\dot{x}_{\mu}\dot{x}_{\nu} K^{\mu\nu}-(L-a\,E)^2\,.
 \end{eqnarray}
The spacetime under consideration admits a non trivial Killing tensor,
\begin{eqnarray}
K^{\mu\nu}&=&\Delta(r)l^{\mu}{n}^{\nu}+r^2g^{\mu\nu}\,,
\end{eqnarray}
where the tetrads $l^{\mu}$ and $n^{\mu}$ mentioned below.
\begin{align}
    \begin{split}
& {l}^{\mu}=\frac{1}{\Delta(r)}\Big[(r^2+a^2+l^2)\delta^{\mu}_{t}+\Delta(r)\delta^{\mu}_{r}+a\delta^{\mu}_{\phi}\Big]\,,\\&
{n}^{\mu}=\frac{1}{\Delta(r)}\Big[(r^2+a^2+l^2)\delta^{\mu}_{t}-\Delta(r)\delta^{\mu}_{r}+a\delta^{\mu}_{\phi}\Big]\,.
\end{split}
\end{align}

\begin{figure}[ht!]
 \centering
 \includegraphics[scale=0.96]{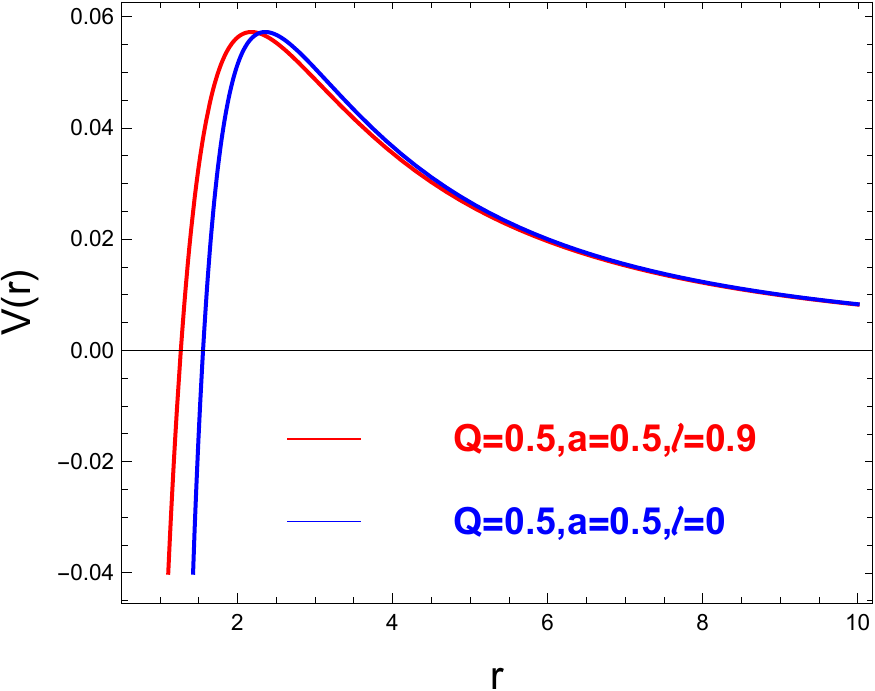}
 \caption{The red and blue curves represent the effective potential $V(r)$ for Kerr-Newman black-bounce and Kerr-Newman spacetime respectively. For large values $l,$ we can see that the maxima of $V(r)$ shifts towards left i.e the radius of photon sphere ($r_c$) becomes smaller.}
\label{Fig1}
\end{figure}

Using these three constants of motion and a convenient choice of parameter the null geodesic equations can be described by the following four first order differential equation,
\begin{align}
    \begin{split}
   & \rho^4 \dot{r}^2=((r^2+l^2+a^2)E-aL)^2-\Delta(r)((L-a\, E)^2+\mathcal{K}):=R^2(r)\,,\\&
    \rho^2 \dot{\phi}=-\Big(a\, E-\frac{L}{\sin^2\theta}\Big)+\frac{a[E(r^2+a^2+l^2)-a\, L]}{\Delta}\,,\\&
    \rho^2 \dot{t}=-a\,(a\, E\,\sin^2\theta-L)+\frac{(r^2+a^2+l^2)[E(r^2+a^2+l^2)-a\, L]}{\Delta}\,,\\&
    \rho^4 \dot{\theta}^2=\mathcal{K}+\cos^2\theta\Big(a^2\, E^2-\frac{L^2}{\sin^2\theta}\Big):=\Theta(\theta)^2\,.\label{211}
\end{split}
\end{align}
In (\ref{211}) the dot stands for the derivative with respect to $\tilde \lambda$. Subsequently, for simplicity we restrict ourselves to the equatorial plane i.e, $\theta=\frac{\pi}{2},\dot{\theta}=0$ and $\mathcal{K}=0.$ Now we can rewrite the radial geodesic equation as,
\begin{eqnarray}
   \frac{\dot{r}^2}{L^2}+V(r)=\frac{1}{\lambda^2}\,, \lambda=\frac{L}{E}\label{212}
\end{eqnarray}
where the effective potential $V(r)$ is defined as,
\begin{eqnarray}
   V(r)=\frac{1}{r^2+l^2}\Big[1-\frac{a^2}{\lambda^2}+\Big(1-\frac{a}{\lambda}\Big)^2\Big(-\frac{2m}{\sqrt{r^2+l^2}}+\frac{Q^2}{r^2+l^2}\Big)\Big]\,.
\label{213}
\end{eqnarray}
$\lambda$ is the impact parameter. One can get a clear idea about the nature of the potential from the Fig.~\ref{Fig1}.
The equation (\ref{212}) is mimicking the motion of a particle in the effective potential $V(r)$. By taking appropriate limits (\ref{212}) reduces to corresponding well-known equations for the Kerr ($l=0, Q=0$) and Schwarzschild black hole ($l=Q=a=0$) respectively. Note that, for Kerr-Newman black hole  ($l=0$), the charge $Q$ gives a repulsive effect to the light rays. This repulsive effect prevents the light rays fall into the black hole. However, for this black-bounce metric, the regularisation parameter $l$ provides an extra attractive effect to the light rays.\par 

Now consider  light rays that start from infinity and approach to the black hole, and then turn back to the infinity to reach the observer. These light rays will have radial turning point namely, the closest approach to the black hole $r_0$, determined by,
\begin{eqnarray} \label{eqq1}
   \Big(\frac{\dot{r}^2}{L^2}\Big)\Big |_{r=r_0}=\frac{1}{\lambda^2}-V(r_0)=0.\label{220}
\end{eqnarray}
From (\ref{eqq1}) we get,
\begin{eqnarray}
     \tilde{r_0}^4-\lambda^2\Big(1-\frac{a^2}{\lambda^2}\Big)\tilde{r_0}^2+2\,m\lambda^2\Big(1-\frac{a}{\lambda}\Big)^2\tilde{r_0}=Q^2\lambda^2\Big(1-\frac{a}{\lambda}\Big)^2 \label{221}
\end{eqnarray}
where, $\tilde{r_0}=\sqrt{r_0^2+l^2}$.\\
Then solving (\ref{eqq1}) we get,
\begin{align}
    \begin{split}
   &  r_0(\lambda)=\Big(\frac{\lambda}{\sqrt{6}}\sqrt{1-\omega^2}\Big[\sqrt{1+\hat \gamma}+\sqrt{2-\hat \gamma-\frac{3\sqrt{6}m(1-\omega)^2}{\lambda(1-\omega^2)^{3/2}\sqrt{1+\hat \gamma}}}\Big]^2-l^2\Big)^{1/2}\label{222}
     \end{split}
\end{align}
where $\omega,\chi,\rho,\hat{\gamma}$ are given by,
\begin{align}
    \begin{split}
  & \omega=\frac{a}{\lambda},\,\, \rho=\frac{12Q^2}{\lambda^2(1+\omega)^2}\,\,,\hat{\gamma} =\sqrt{1-\rho}\cos\Big(\frac{2\chi}{3}\Big), \\&
     \chi=\arccos\Big(\frac{3\sqrt{3}\,m(1-\omega)^2}{\lambda(1-\omega^2)^{3/2}(1-\rho)^{3/4}}\sqrt{1-\frac{\lambda^2(1+\omega)^3}{54\,m^2(1-\omega)}\Big[1+3\rho-(1-\rho)^{3/2}\Big]}\Big)\,.\label{223}
     \end{split}
\end{align}
Eq.~(\ref{222}) reduces to that of Kerr-Newmann for $l=0.$ 
\begin{figure}[ht!]
\centering
\includegraphics[scale=0.55]{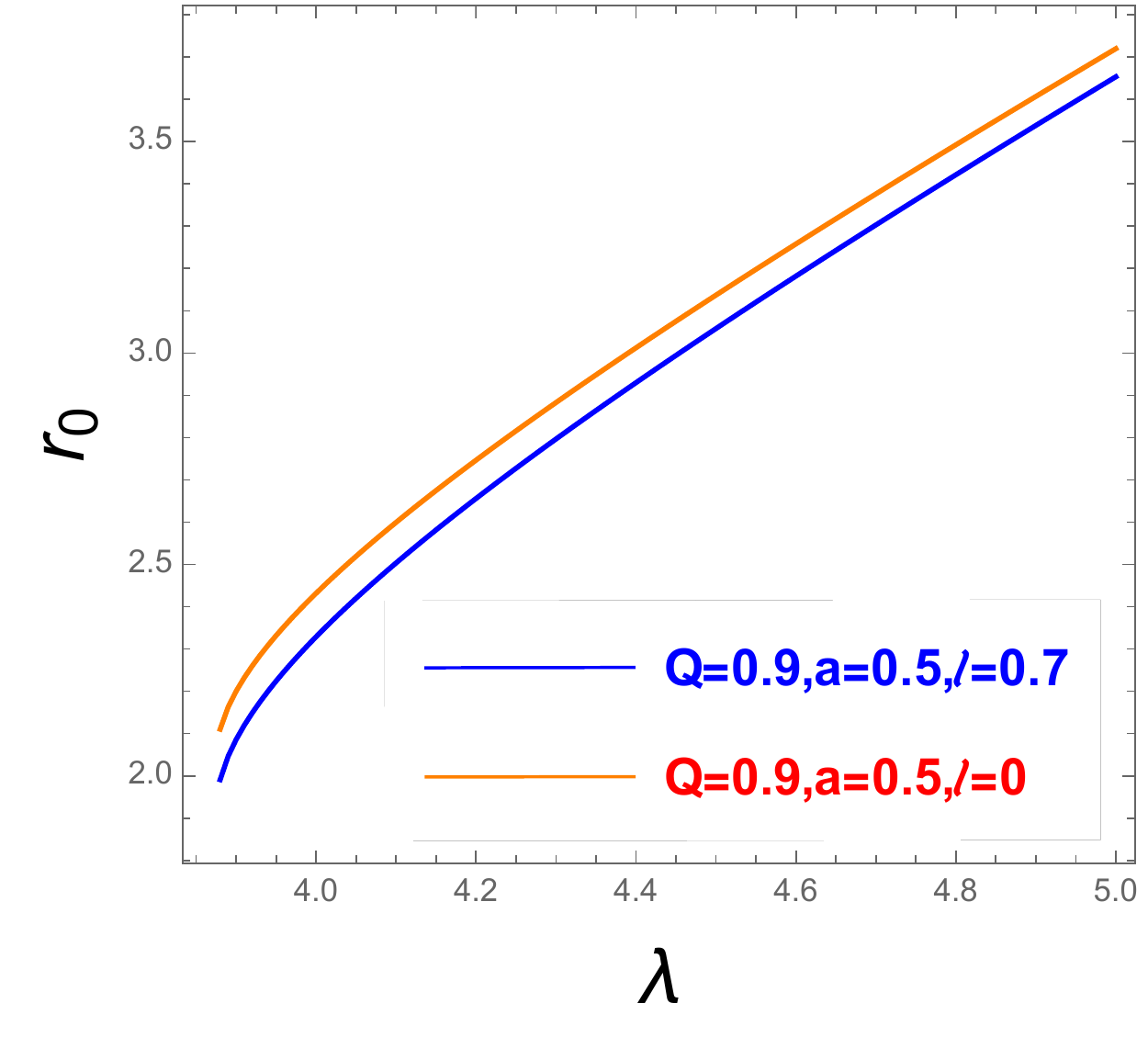}
\includegraphics[scale=0.55]{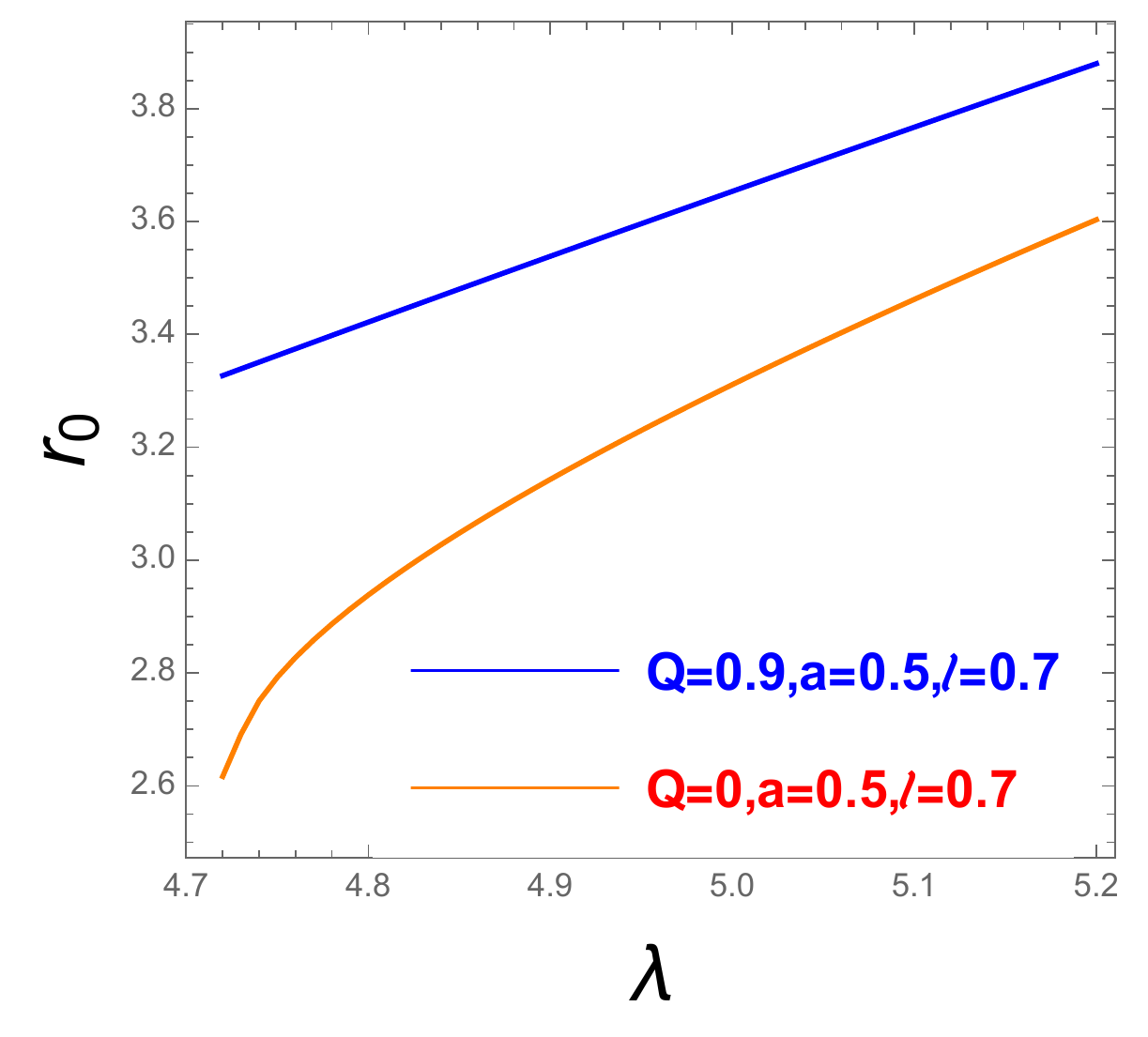}
\caption{Variation of the distance of closest approach $r_0$ with respect to the impact parameter $\lambda$ for different choices of black hole parameters. The left one shows the variation of $r_0$ with respect to $\lambda$ for different regularisation parameters ($l$). The right one shows the variation of $r_0$ with respect to $\lambda$ for different charge parameters ($Q$).}
\label{Fig:2}
\end{figure}
As evident from Fig.~(\ref{Fig:2}), the additional attractive effect because of $l$ pulls the distance of closest approach $r_0.$ As can be seen from Fig.~(\ref{Fig:2}), for a fixed impact parameter $\lambda$, $r_0$ is smaller for the rotating charged black-bounce metric compared to that of its Kerr-Newman counterpart. From Fig. (\ref{Fig:2}), it is also evident that the non zero charge $Q$ gives an additional repulsive effect on the light rays.   \\\\
\textit{{\bf Critical impact parameter and the photon sphere:}}\\\\
We end this section by discussing the position of the photon sphere and the critical impact parameter for the light rays coming from asymptotic infinity. This will be useful for the analysis of deflection in the subsequent sections. From Fig.(\ref{Fig1}) we can see that the effective potential $V(r)$ in (\ref{213}) vanishes at infinity and \textit{attain maximum value at the  $r=r_c.$} This critical radius $r_c$ is the radius of the radius of photon sphere. This is defined by the following condition, 
\begin{eqnarray}
   \frac{\partial V(r)}{\partial r}\Big|_{r=r_c}=0\,. \label{vmax}
\end{eqnarray}
From (\ref{vmax}) we can find out 
\begin{eqnarray}
     r_c(\lambda_c)=\sqrt{\Big(\frac{3\,m\,\hat \zeta}{2}\Big[1+\sqrt{1-\zeta}\Big]\Big)^2-l^2\Big)}\label{33}
\end{eqnarray}
where, $$\tilde{r}=\sqrt{r^2+l^2},\quad \hat \zeta= \Big(\frac{1-\frac{a}{\lambda_c}}{1+\frac{a}{\lambda_c}}\Big), \quad \zeta=\frac{8\,Q^2}{9\,m^2}\hat \zeta\,. $$  One can easily check that the result presented in (\ref{33}) can be reduced to that of Kerr-Newmann black holes black hole in limit $l=0$. \par

Note that the turning point of the photon is $r_0$ defined in (\ref{222}) attains its minimum value at $r_c$. The corresponding value of the impact parameter is defined as $\lambda_c.$ If for some $\lambda,$ $r_0$ becomes less than $r_c$ then the photon will be captured by the black hole. Hence $\lambda_c$ is known as the critical impact parameter. Finally, substituting equation (\ref{222}) into (\ref{33}) we can write  $\lambda_c=\lambda_c(a,Q,l)\,\, \textrm{and} \,\, r_c=r_c(a,Q,l).$\par

\section{Computation of deflection angle: Analytical approach} \label{sec3}
 The study of gravitational lensing has been very important in recent days because it has a direct consequence on the observational appearances of the black holes \cite{EventHorizonTelescope:2019dse}. In this section, we will present some analytical results for deflection angle for photon passing near a Kerr-Newman black-bounce spacetime. This is one of the main goals of this paper. The first task is to calculate the deflection angle of the photon in some preferred plane. For simplicity we choose the equatorial plane, i.e, $\theta=\frac{\pi}{2}$ and $\dot{\theta}=0$. In order to study the lensing of light rays by a black hole spacetime, we need to take a strong deflection limit, i.e. when the turning point of photon $r_0$ is very close to the radius of photon sphere $r_c.$ We can do it in two different ways: first, calculate the exact deflection angle and take the strong deflection limit or take the strong deflection limit from the beginning and arrive at the formula for the deflection angle. In this section, we give an analytical calculation of the exact deflection angle in perturbative order. In the next section, we adopt the second approach, calculate the deflection angle in the strong deflection limit, and discuss its observational consequences. \par 
 To calculate the deflection angle we will follow the method of \cite{Iyer:2009wa, Hsiao:2019ohy}.
Before proceeding with the computation we define the following coordinate, 
\begin{eqnarray}
   u:=\frac{1}{\sqrt{r^2+l^2}}\,.
\end{eqnarray}
Then combining the first two equations in (\ref{211}) we get,
\begin{eqnarray}
   \Big(\frac{du}{d\phi}\Big)^2=\Big(\frac{du}{dr}\frac{dr}{d\phi}\Big)^2=\frac{r^2}{(r^2+l^2)^3}\frac{\dot{r}^2}{\dot{\phi}^2}=(u^4-l^2 u^6)\frac{\dot{r}^2}{\dot{\phi}^2}\,.
\label{42}\end{eqnarray}
Now our goal is to rewrite the $\dot{r}$ and $\dot{\phi}$ as function of $u$.
We have,
\begin{eqnarray}
    \dot{r}^2&=&L^2\Big(\frac{1}{\lambda^2}-V(r)\Big)\nonumber\\&=&L^2\Bigg[\frac{1}{\lambda^2}-\Big[\frac{1}{r^2+l^2}\Big(1-\frac{a^2}{\lambda^2}+(1-\frac{a}{\lambda})^2(-\frac{2m}{\sqrt{r^2+l^2}}+\frac{Q^2}{r^2+l^2}\Big)\Big]\Bigg]\nonumber\\&=&L^2\Big[\frac{1}{\lambda^2}-u^2\Big(1-\frac{a^2}{\lambda^2}\big)-Q^2\Big(1-\frac{a}{\lambda}\Big)^2 u^4+2m\Big(1-\frac{a}{\lambda}\Big)^2 u^3\Big]:=L^2 B(u)\,.\label{43}
\end{eqnarray}
Now mimicking the same procedure we can write,
\begin{eqnarray}
   \dot{\phi}^2&=&\frac{1}{(r^2+l^2)^2}\Big[-(a\, E-L)+\frac{a[(E(r^2+a^2+l^2))-aL]}{r^2+a^2+l^2-2m\sqrt{r^2+l^2}+Q^2}\Big]^2\nonumber\\&=&\frac{1}{u^4}\Big[E\,\lambda\,\Big(1-\frac{a}{\lambda}\Big)+\frac{a\,E(1+a^2\,u^2-a\,u^2\lambda)}{1+a^2u^2-2mu+Q^2u^2}\Big]^2\nonumber\\&=&\frac{L^2}{u^4}\Bigg[\frac{1-(2\,m\,u-Q^2u^2)(1-\frac{a}{\lambda})}{1-2\,m\,u+(a^2+Q^2)u^2}\Big]^2\,.\label{44}
\end{eqnarray}
Combining equation (\ref{43}) and (\ref{44}) and using the equation (\ref{42}) we  get,
\begin{eqnarray}
   \Big(\frac{du}{d\phi}\Big)^2=(1-l^2u^2)\Bigg[\frac{1-2mu+(a^2+Q^2)u^2}{1-(2mu-Q^2u^2)(1-\frac{a}{\lambda})}\Bigg]^2 B(u)\,.\label{45}
\end{eqnarray} 
The deflection angle of the photon $\hat{\alpha}$ can be calculated by integrating the equation (\ref{45}) over $u$ from $0$ to $\frac{1}{\sqrt{r_0^2+l^2}}$, where $r_0$ is the turning point and then  evaluating the resulting expression at the critical value $r_c$ \cite{Bozza:2002af}, 
\begin{eqnarray}
   \hat{\alpha}=-\pi+2\int_0 ^\frac{1}{\sqrt{r_0^2+l^2}}\frac{1}{\sqrt{1-l^2u^2}}\Bigg[\frac{1-(2mu-Q^2u^2)(1-\frac{a}{\lambda})}{1-2mu+(a^2+Q^2)u^2}\Bigg]\frac{1}{\sqrt{B(u)}}\,.\label{46}
\end{eqnarray}
In (\ref{46}), the polynomial $B(u)(1-l^2u^2)$ is of \textit{degree six}. Hence we can not write this integral as a complete form of elliptic integral. However to make some progress analytically we will assume 
    \begin{equation}l^2u^2<<1.
    \end{equation}
    Then we can Taylor expand $\frac{1}{\sqrt{1-l^2u^2}}=1+\frac{l^2u^2}{2}+\mathcal{O}(l^4u^4)$. Therefore we can rewrite (\ref{46}) upto $\mathcal{O}(l^4)$,
\begin{eqnarray}
   \hat{\alpha}&=&-\pi+2\int_0 ^\frac{1}{\sqrt{r_0^2+l^2}}\,du\,\Big(1+\frac{l^2u^2}{2}\Big)\Big[\frac{1-(2\,m\,u-Q^2u^2)(1-\omega)}{1-2\,m\,u+(a^2+Q^2)u^2}\Big]\frac{1}{\sqrt{B(u)}}+\mathcal{O}(l^4)\nonumber \\&=&\hat{\alpha}_{\textrm{KN}}+l^2*\text{(correction term)}+\mathcal{O}(l^4)\nonumber \\&=&\hat{\alpha}_{KN}\Big|_{\sqrt{{r_0^2+l^2}}}+l^2*\boldsymbol\xi(m,a,\lambda,Q)+\mathcal{O}(l^4)\,.\label{eqq2}
\end{eqnarray}
$\hat \alpha_{\textrm{KN}}$ corresponds to the deflection angle for Kerr-Newman black hole which can be obtained from (\ref{eqq2}) in $l=0$ limit and 
\begin{eqnarray}
   \boldsymbol{\xi}(m,a,\lambda,Q)&=& \int_0^{\frac{1}{\sqrt{r_0^2+l^2}}} u^2 \Bigg[\frac{1-(2mu-Q^2u^2)(1-\omega)}{1-2mu+(a^2+Q^2)u^2}\Bigg]\frac{1}{\sqrt{B(u)}}\,du\,,\label{412}
\end{eqnarray}
with $\omega=\frac{a}{\lambda}.$ We now rewrite $B(u)$ as,
\begin{eqnarray}
   B(u)= - Q^2(1-w)^2(u-u_1)(u-u_2)(u-u_3)(u-u_4)\,,\label{310}
\end{eqnarray}
where the roots are defined as,
\begin{eqnarray}
 u_1&=&\frac{X_1-2m-X_2}{4m\sqrt{r_0^2+l^2}}\,,\label{311}\\
 u_2&=&\frac{1}{\sqrt{r_0^2+l^2}}\,,\\
 u_3&=&\frac{X_1-2m+X_2}{4m\sqrt{r_0^2+l^2}}\,,\\
 u_4&=&\frac{2m}{Q^2}-\frac{X_1}{2m\sqrt{r_0^2+l^2}}\,.\label{314}
\end{eqnarray}
By suitably choosing the constants $X_1$ and $X_2$ allows us to write down the roots in the following order $u_1<u_2<u_3<u_4$. Among them $u_2,u_3,u_4$ are positive roots while $u_1$ is a negative root. In order to extract the roots, we need to substitute equations (\ref{311}) to (\ref{314}) into (\ref{310}) and comparing it with the coefficients of $u^0,u^2,u^3,u^4$ in (\ref{43}) we will get \cite{Hsiao:2019ohy},
\begin{align}
\begin{split}
     & Q^2\Big[X_2^2-(X_1-2m)(X_1+6m)+4X_{1}^2\Big]=16m^2\sqrt{r_0^2+l^2}\Big(X_1-\sqrt{r_0^2+l^2}\frac{1+\omega}{1-\omega}\Big)\,,\\ &
     X_2^2-(X_1^2-2m)^2=\frac{8m(X_1-2m)(Q^2X_1-4m^2\sqrt{r_0^2+l^2})}{Q^2(X_1-2m)-4m^2\sqrt{r_0^2+l^2}}\,,\\ &
     \Big[X_2^2-(X_1^2-2m)^2\Big]\Bigg(\frac{1}{8m(r_0^2+l^2)^{\frac{3}{2}}}-\frac{Q^2X_1}{32m^3(r_0^2+l^2)^2}\Bigg)=\frac{1}{\lambda^2(1-\omega)^2}\,.\label{418}
     \end{split}
\end{align}
Combining the first and second equation of (\ref{418}) gives the equation for $X_1$, which is given by,
\begin{align}
\begin{split}
   &  \frac{Q^2}{2m}X_1^3-\Big(Q^2+4m\sqrt{r_0^2+l^2}\Big)X_1^2+\Bigg(4m^2\sqrt{r_0^2+l^2}+2mQ^2+\frac{8m^3(r_0^2+l^2)}{Q^2}+\frac{2m((r_0^2+l^2)(1+\omega)}{(1-\omega)}\Bigg)X_1\\ &
   =4\,m^2\,Q^2+\frac{4m^2(r_0^2+l^2)(1+\omega)}{(1-\omega)}+\frac{8m^2(r_0^2+l^2)^{\frac{3}{2}}(1+\omega)}{Q^2(1-\omega)}\,.\label{419}
\end{split}
\end{align}
We can analytically solve the equation(\ref{419}) and the positive real root is given by,
\begin{align}
\begin{split}
& \textstyle{X_1(m,Q,\omega,l,r_0)=\frac{2m(Q^2+4m\sqrt{r_0^2+l^2})}{3Q^2}
+\frac{8m^2\sqrt{r_0^2+l^2}}{3Q^2}\sqrt{1+\frac{Q^2}{2m^2}\Big(\frac{m}{\sqrt{r_0^2+l^2}}-\frac{3(1+\omega)}{2(1-\omega)}-\frac{Q^2}{r_0^2+l^2}}\Big)\,      \cos\Big(\frac{\delta}{3}+\frac{2\pi}{3}\Big)}\,.
\end{split}
\end{align}
where,
\begin{align}
\begin{split}
   & \textstyle{\delta\,\, =\,\, \arccos\Big(\frac{-8m^3(r_0^2+l^2)^{3/2}-3\,m\,Q^2(r_0^2+l^2)\Big(2m-\frac{3\sqrt{r_0^2+l^2}(1+\omega)}{(1-\omega)}\Big)-3\,Q^4\sqrt{r_0^2+l^2}\Big(5m-\frac{3\sqrt{r_0^2+l^2}(1+\omega)}{(1-\omega)}\Big)+10\,Q^6}{\Big[4m^2(r_0^2+l^2)+Q^2\sqrt{r_0^2+l^2}\Big(2m-\frac{3\sqrt{r_0^2+l^2}(1+\omega)}{(1-\omega)}\Big)-2\,Q^4\Big]^{3/2}}\Big)}\,.\label{equ21}
\end{split}
\end{align}
At this point, we can check that in the limit $Q=0, l=0, $ $\delta$ becomes $\pi.$ Then (\ref{equ21}) can be written as,
\begin{eqnarray}
    X_1(m,\omega,Q=0,l=0,r_0)=r_0\frac{1+\omega}{1-\omega}\,.
\end{eqnarray}
This is exactly the result for Kerr black hole \cite{Iyer:2009wa}.
For Reissner-Nordstrom black hole we have $a=0, l=0$ and the roots of the equation (\ref{419}) then reduces to,
\begin{equation}
    X_1(m,Q,\omega=1,l=0,r_c)=2m\Big(\frac{2mr_0}{Q^2}-1\Big)\Big|_{r_c}\label{320}
\end{equation}
where $r_c$ is defined in (\ref{33})
and the result (\ref{320}) matches with the known result \cite{Hsiao:2019ohy}.\par
Now we turn our attention to the function mentioned inside the third bracket of (\ref{412}). Let
\begin{eqnarray}
 F(u)= 1-2\,m\,u+(a^2+Q^2)u^2=(u-u_-)(u-u_+)\,,\label{420}
\end{eqnarray}

where $u_{-}$ and $u_{+}$ are the roots of the equation $F(u)$. Therefore we can rewrite the following term in the manner described below.
\begin{eqnarray}
\displaystyle{\frac{1-2mu(1-\omega)+Q^2u^2(1-\omega)}{1-2mu+(a^2+Q^2)u^2} =\frac{K_+}{u_+-u}+\frac{K_-}{u_--u}+\frac{K_{Q+}\,u}{u_+-u}+\frac{K_{Q-}\,u}{u_- -u}}\label{equ25}
\end{eqnarray}
where,
\begin{eqnarray} \label{upos}
 u_{\pm}=\frac{m\pm\sqrt{m^2-(a^2+Q^2)}}{a^2+Q^2}\,.
\end{eqnarray}
As evident from (\ref{upos}) $u_\pm$ are positive.
Solving equation (\ref{equ25}), we can find out the constants $K_-,K_+,K_{Q+},K_{Q-}$ and given by,
\begin{eqnarray}
     K_+&=&\frac{2m(1-\omega)\Big(m+\sqrt{m^2-(a^2+Q^2)}\Big)-(a^2+Q^2)}{2(a^2+Q^2)\sqrt{m^2-(a^2+Q^2)}},\\
     K_-&=&\frac{(a^2+Q^2)-2m(1-\omega)\Big(m-\sqrt{m^2-(a^2+Q^2)\Big)}}{2(a^2+Q^2)\sqrt{m^2-(a^2+Q^2)}},\\
     K_{Q+}&=&\frac{-Q^2(1-\omega)\Big(m+\sqrt{m^2-(a^2+Q^2)}\Big)}{2(a^2+Q^2)\sqrt{m^2-(a^2+Q^2)}},\\
     K_{Q-}&=& \frac{Q^2(1-\omega)\Big(m-\sqrt{m^2-(a^2+Q^2)}\Big)}{2(a^2+Q^2)\sqrt{m^2-(a^2+Q^2)}}.
\end{eqnarray}
Finally integral in (\ref{412}) can be written as,\\
\begin{align}
\begin{split}
  \textstyle{\boldsymbol{\xi}(m,a,\lambda,Q)} & =
\textstyle{ \int_0^{u_2} du\,  u^2\Big[\frac{K_+}{u_+-u}+\frac{K_-}{u_--u}+\frac{K_{Q+}u}{u_+-u}+\frac{K_{Q-}u}{u_--u}\Big]\frac{1}{\sqrt{-Q^2(1-\omega)^2(u-u_1)(u-u_2)(u-u_3)(u-u_4)}}}\\ &
\textstyle{ = K_{+}\,g\, (u_+\,\Delta\,\textbf{F}+[(u_1-u_4)\Delta\boldsymbol{\Pi}(\alpha^2)+u_4\Delta\,\textbf{F}]
     -u_+^2\frac{1}{(u_+-u_1)}[\frac{u_1-u_4}{u_+-u_4}\Delta\boldsymbol{\Pi}(\alpha_{+3}^2)+\frac{u_+-u_1}{u_+-u_4}\Delta\textbf{F}])}\\ &
   \textstyle{+K_{-}\,g\,(u_-\,\Delta\textbf{F}+[(u_1-u_4)\Delta\boldsymbol{\Pi}(\alpha^2))+u_4\Delta\textbf{F}]
     -u_{-}^2\frac{1}{(u_--u_1)}[\frac{u_1-u_4}{u_--u_4}\Delta\boldsymbol{\Pi}(\alpha_{-3}^2)+\frac{u_--u_1}{u_--u_4}\Delta\textbf{F}])}\\ &
   \textstyle{+K_{Q+}\, g [(u_4^2-2u_+u_4+u_+^2-\frac{u_+^3}{u_+-u_4})\Delta\textbf{F}+(u_+(u_1-u_4)-\frac{u_+^3(u_1-u_4)}{(u_+-u_1)(u_+-u_4)})\Delta\boldsymbol{\Pi}(\alpha_{+3}^2)}\\ & \textstyle{-2u_4(u_1-u_4)\boldsymbol{V_1}(\alpha^2)-(u_1-u_4)^2\boldsymbol{V_2}(\alpha^2)]+K_{Q-}\, g\,[(u_4^2-2u_-u_4+u_-^2-\frac{u_-^3}{u_--u_4}) \Delta\textbf{F}}\\ & \textstyle{+(u_-(u_1-u_4)-\frac{u_-^3(u_1-u_4)}{(u_--u_1)(u_--u_4)}) 
\Delta\boldsymbol{\Pi}(\alpha_{-3}^2)-2u_4(u_1-u_4)\boldsymbol{V_1}(\alpha^2)-(u_1-u_4)^2\boldsymbol{V_2}(\alpha^2)]}\label{431}
 \end{split}
 \end{align}
where,
\begin{align}
    \begin{split}
   & \Delta\,\textbf{F}=F(\phi,k)-F(\frac{\pi}{2},k),
   \,\,\Delta\boldsymbol{\Pi}(\alpha^2)=\Pi(\phi,\alpha^2,k)-\Pi(\frac{\pi}{2},\alpha^2,k),\,\,\\&\Delta\boldsymbol{\Pi}(\alpha_{\pm3}^2)=\Pi(\phi,\alpha_{\pm3}^2,k)-\Pi(\frac{\pi}{2},\alpha_{\pm3}^2,k)\label{432}
    \end{split}
 \end{align}
 and, $\boldsymbol{V_1}$ and $\boldsymbol{V_2}$ is given by,
\begin{align}
\begin{split}
& \boldsymbol{V_1}(\alpha^2)=\Pi(\frac{\pi}{2},\alpha^2,k)-\Pi(\phi,\alpha^2,k)\,,\\ &
\boldsymbol{V_2}(\alpha^2)=\frac{1}{2(\alpha^2-1)(k^2-\alpha^2)}\Big[\alpha^2\Big(E(\frac{\pi}{2},k)-E(\phi,k)\Big)+(k^2-\alpha^2)\Big(F(\frac{\pi}{2},k)-F(\phi,k)\Big)\,,\\ &
(2\alpha^2\,k^2+2\alpha^2-\alpha^4-3k^2)\Big(\Pi(\frac{\pi}{2},\alpha^2,k)\Big)-\Pi(\phi,\alpha^2,k)\Big)\Big)-\frac{\alpha^4\,\sin\,\phi\sqrt{1-\sin^2\phi}\sqrt{1-k^2\,\sin^2\phi}}{1-\alpha^2 \,\sin^2\phi}\Big]
\end{split}
\end{align}
 and 
 \begin{align}
     \begin{split}
           & g=\frac{2}{\sqrt{Q^2(1-\omega)^2(u_4-u_2)(u_3-u_1)}},\,\, \alpha^2=\frac{u_1-u_2}{u_4-u_2}<0,\,\,
     \,\,
      \alpha_{\pm 3}^2=\alpha^2\frac{u_{\pm}-u_4}{u_{\pm}-u_1}\,,\\&
       k^2=\frac{(u_4-u_3)(u_2-u_1)}{(u_4-u_2)(u_3-u_1)},\,\,
       \phi=\arcsin\sqrt{\frac{(X_2+2m-X_1)[4m^2\sqrt{r_0^2+l^2}-Q^2(X_1+2m)]}{(X_2+6m-X_1)(4m^2\sqrt{r_0^2+l^2}-Q^2X_1)}}.
     \end{split}
 \end{align}
Note that, $\Pi(\phi,\alpha^2,k)$ and $\Pi(\frac{\pi}{2},\alpha^2,k)$ are the incomplete and complete elliptic integral of third kind respectively. Also, $F(\phi,k)$ and $F(\frac{\pi}{2},k)$ are the incomplete and complete elliptic integral of first kind. Additionally, in $\boldsymbol{V}_{1,2}$ we have $E(\phi,k)$ and $E(\frac{\pi}{2},k)$, which are incomplete and complete elliptic integral of second kind respectively. The details of the computation are given in the Appendix ~(\ref{App}).\par
Together with (\ref{eqq2}), (\ref{431}) provides an expression of equatorial deflection angle for Kerr-Newman black-bounce metric upto $\mathcal{O}(l^2).$  This result is one of the main results of the paper.

\section{Strong deflection analysis of equatorial lensing}\label{sec4}
To get further insight into the deflection angle for our context and make contact with the possible observational signature, in this section, we study the strong deflection limit \cite{Bozza:2002af} of the equatorial deflection angle ($\theta=\frac{\pi}{2}$) of the light rays mentioned in the previous section.  It is difficult to take the strong deflection limit of (\ref{431}) directly. It will be easier to take the strong field limit of the integrand of (\ref{46}) first and then do the integration. As mentioned before, we will only consider those light rays whose turning point is very close to the radius of photon sphere. \par
On the equatorial plane the metric takes the following form,
\begin{eqnarray}
   ds^2=-\mathcal{A}(r)dt^2+\mathcal{B}(r)dr^2+\mathcal{C}(r)d\phi^2-\mathcal{D}(r)dt\,d\phi
\end{eqnarray}
with
\begin{align}\begin{split}
   &\mathcal{A}(r)=\frac{1}{\rho^2}(\Delta(r)-a^2)\,,\,\,\, \mathcal{B}(r)=\frac{\rho^2}{\Delta}\,,
   \mathcal{C}(r)=\frac{1}{\rho^2}\Big[(r^2+a^2+l^2)^2-\Delta(r)a^2\Big]\,,\\&
   \mathcal{D}(r)=\frac{2}{\rho^2}\Big[(r^2+a^2+l^2)a-\Delta(r)a\Big]\,.
\end{split}\end{align}Note that all metric components are evaluated at equatorial plane. We have already seen that due to the existing symmetries, the spacetime admits two conserved quantities $E$ and $L$. For the sake of simplicity we set $E=1$. Therefore the impact parameter $\lambda=L$. From (\ref{212}), after using the fact that at the distance of closest approach $r=r_0$ and $\dot{r}=0,$ we get the following, 
\begin{align}
\begin{split}
  & L=\frac{-\mathcal{D}_0+\sqrt{4\mathcal{A}_0\mathcal{C}_0+\mathcal{D}_0^2}}{2\mathcal{A}_0}\\ &
  =\frac{\sqrt{l^2+r_0^2} \left(\left(l^2+r_0^2\right) \sqrt{a^2-2\,m\, \sqrt{l^2+r_0^2}+l^2+Q^2+r_0^2}+a \left(Q^2-2\,m\, \sqrt{l^2+r_0^2}\right)\right)}{Q^2 \sqrt{l^2+r_0^2}+l^2 \left(\sqrt{l^2+r_0^2}-2\,m\,\right)+r_0^2 \left(\sqrt{l^2+r_0^2}-2\,m\,\right)}\,.
\end{split} \label{433}
\end{align}
The subscript $``0"$ denotes functions are evaluated at $r=r_0$. From the equation of motion of $\phi$ (the second equation of (\ref{211})) we get,
\begin{eqnarray}\label{az}
   \phi(r_0)=2\int_{r_{0}}^\infty\frac{\sqrt{\mathcal{B}\mathcal{A}_0}(\mathcal{D}+2L\mathcal{A})}{\sqrt{4\mathcal{AC}+\mathcal{D}^2}\sqrt{\mathcal{CA}_0-\mathcal{AC}_0+L(\mathcal{AD}_0-\mathcal{DA}_0)}}\,dr\,.
\end{eqnarray}
In the strong field limit, we consider only those photons whose distance of closest approach $r_0\approx r_c,$ where $r_c$ is the radius of the photon sphere. Hence the deflection angle $\alpha$ (\ref{46}) can be expanded in terms of $r_c$ or equivalently the critical impact parameter $\lambda_c$. When the distance of the closest approach $r_0$ is greater than $r_c$, the light rays get deflected (but it can orbit around the black hole several times before reaching the observer). Otherwise, around $r=r_c$, $\hat{\alpha}$ diverges and the photon gets absorbed by the black hole. Following the method developed here \cite{Bozza:2002zj,Bozza:2002af}, one can find out the nature of the divergence in the deflection angle when the photons are very close to the radius of photon sphere $r=r_c$. One can define two variables $y,z$ as,
\begin{eqnarray}
   y=\mathcal{A}(r),
   z=\frac{y-y_0}{1-y_0}\,.
\end{eqnarray}
Now one can express azimuthal angle defined in (\ref{az}) in terms of these two new variables,
\begin{eqnarray}
   \phi(r_0)=\int_0^1\mathcal{R}(z,r_0)\mathcal{F}(z,r_0) dz\label{59}
\end{eqnarray}
where,
\begin{eqnarray}
   \mathcal{R}(z,r_0)&=&\frac{2(1-y_0)}{A'}\frac{\sqrt{\mathcal{B}\mathcal{A}_0}(\mathcal{D}+2L\mathcal{A})}{\sqrt{4\mathcal{AC}^2+\mathcal{CD}^2}}\\
   \mathcal{F}(z,r_0)&=&\frac{1}{\sqrt{\frac{1}{\mathcal{C}}(\mathcal{CA}_0-\mathcal{AC}_0+L(\mathcal{AD}_0-\mathcal{DA}_0)})}=\frac{1}{\sqrt{\mathcal{H}}}\,,\label{H1}\\
   \mathcal{H} & =&\frac{1}{\mathcal{C}}(\mathcal{CA}_0-\mathcal{AC}_0+L(\mathcal{AD}_0-\mathcal{DA}_0))\,.\label{H}
\end{eqnarray}
One can see that the function $\mathcal{R}(z,r_0)$ is regular for any values of $z$ and $r_0$ but the function $\mathcal{F}(z,r_0)$ is divergent for $z=0$, i.e at $r=r_0$. Therefore one can rewrite (\ref{az}) after extracting the divergent part,
\begin{eqnarray}
   \phi(r_0)=\phi_{\mathcal{R}}(r_0)+\phi_{\mathcal{F}}(r_0)
\end{eqnarray}
where the divergent part can be written as,
\begin{eqnarray}
   \phi_{\mathcal{F}}(r_0)=\int_0^1\mathcal{R}(z=0,r_c)\mathcal{F}_0(z,r_0)dz\,.\label{513}
\end{eqnarray}
As we know, at $r_0=r_c$ the deflection angle should diverge, signalling that the black hole captures the photon. Then our goal is to find out the nature of the divergence. We can find out the nature of the divergence by investigating the denominator (\ref{59}). With the aim of doing so we Taylor expand the denominator of $\mathcal{F}_0(r_0,z)$(\ref{514}) around $z=0$.
\begin{eqnarray}
   \mathcal{F}_{0}(z,r_0)\approx \frac{1}{\sqrt{\gamma_1(r_0)z+\gamma_2(r_0) z^2+\mathcal{O}(z^3})}\,.\label{514}
\end{eqnarray}
Note that, if $\gamma_1(r_0)=0,$ (this happens when $r_0$ coincides with the radius of photon sphere   \cite{Claudel:2000yi})  then it is evident from (\ref{514}), the leading term goes as $\frac{1}{z}$ in small $z$ limit. Hence after the integration will give a logarithmic divergence as shown in (\ref{523}).
To identify $\gamma_1$ and $\gamma_2$ we first Taylor expand $\mathcal{H}$ defined in (\ref{H}).
\begin{eqnarray}
   \mathcal{H}(z,r_0)=\mathcal{H}(0,r_0)+\frac{\partial\mathcal{H}}{\partial z}\Big|_{z=0}z+\frac{1}{2!}\frac{\partial^2\mathcal{H}}{\partial z^2}\Big|_{z=0}z^2+\mathcal{O}(z^3),\text{with}\,\,\mathcal{H}(0,r_0)=0\,.
\end{eqnarray}
Therefore using (\ref{H1}) we can identify $\gamma_1$ and $\gamma_2$ as ,
\begin{eqnarray}
   \gamma_1&:=&\frac{\partial \mathcal{H}}{\partial z}\Big|_{z=0}\nonumber\\
   &=&\frac{1-\mathcal{A}_0}{\mathcal{A}_0'\mathcal{C}_0}\Bigg(\mathcal{A}_0\mathcal{C}_0'-\mathcal{A}_0'\mathcal{C}_0-L\Big(\mathcal{A}_0\mathcal{D}_0'-\mathcal{A}_0'\mathcal{D}_0\Big)\Bigg)
\end{eqnarray}
and
\begin{align}
\begin{split}
  &  \gamma_2:=\frac{1}{2!}\frac{\partial^2 \mathcal{H}}{\partial z^2}\Big|_{z=0}\\ &
   =\frac{(1-\mathcal{A}_0)^2}{2\mathcal{C}_0^2\mathcal{A}_0'^3}\Big[2\mathcal{C}_0\mathcal{C}_0'\mathcal{A}_0'^2+\Big(\mathcal{C}_0\mathcal{C}_0''-2\mathcal{C}_0'^2\Big)\mathcal{A}_0\mathcal{A}_0'-\mathcal{C}_0\mathcal{C}_0'\mathcal{A}_0\mathcal{A}_0''  \\ &
   +L\Big(\mathcal{A}_0\mathcal{C}_0(\mathcal{A}_0''\mathcal{D}_0'-\mathcal{A}_0'\mathcal{D}_0'')+2\mathcal{A}_0'\mathcal{C}_0'(\mathcal{A}_0\mathcal{D}_0'-\mathcal{A}_0'\mathcal{D}_0)\Big)\Big]\,.
\end{split}
\end{align}
The regular part can be written as,
\begin{eqnarray}
   \phi_{\mathcal{R}}(r_0)=\int_0^1\mathcal{G}(z,r_0)dz\label{518}
\end{eqnarray}
where $\mathcal{G}(z,r_0)=\mathcal{R}(z,r_0)\mathcal{F}(z,r_0)-\mathcal{R}(z=0,r_c)\mathcal{F}_0(z,r_0)$.
As we discussed the coefficient $\gamma_1=0.$ This implies, 
\begin{eqnarray}
  \mathcal{A}_0\mathcal{C}_0'-\mathcal{A}_0'\mathcal{C}_0-L\, (\mathcal{A}_0\mathcal{D}_0'-\mathcal{A}_0'\mathcal{D}_0)\Big|_{r_0=r_c}=0\,.\label{519}
\end{eqnarray}
From (\ref{519}) we get,
\begin{align}
    \begin{split}
& \textcolor{black}{2 a^2 \left(Q^2-\sqrt{l^2+r_c^2}\right)+2 a \left(\sqrt{l^2+r_c^2}-Q^2\right) \sqrt{a^2-2 \sqrt{l^2+r_c^2}+l^2+Q^2+r_c^2}+l^4+}\\ &\textcolor{black}{
l^2 \left(-5 \sqrt{l^2+r_c^2}+3 Q^2+2 r_c^2+6\right)-7 Q^2 \sqrt{l^2+r_c^2}-5 r^2 \sqrt{l^2+r_c^2}+2 Q^4+3 \left(Q^2+2\right) r_c^2+r_c^4=0}
    \end{split}\label{r0}
\end{align}

It can be easily checked that, using the expression for critical impact parameter $\lambda_c$ as mentioned in (\ref{433}), one gets the same expression for the radius of photon sphere as mentioned in (\ref{33}). So we can say that (\ref{r0}) gives yet another definition of the photon sphere. For more details on the geometry of photon spheres and various complementary definitions of photon spheres, interested readers are referred \cite{Claudel:2000yi}. \par
\begin{figure}[ht!]
		\minipage{0.33\textwidth}
\includegraphics[scale=0.60]{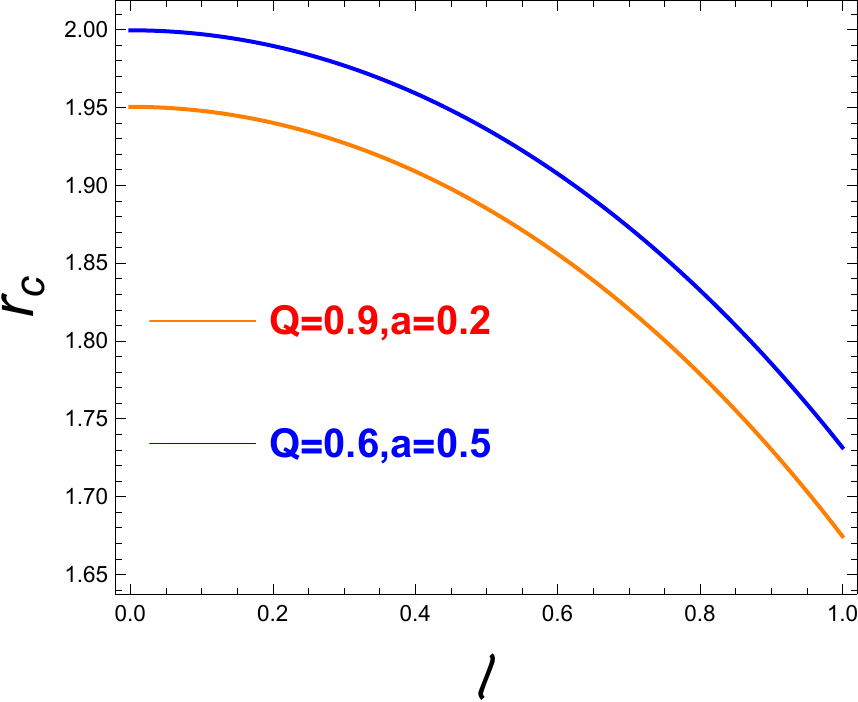}
	\endminipage\hfill
		%%%%%%%%%%%%%%%%%%%%%%%%
		\minipage{0.33\textwidth}
\includegraphics[scale=0.60]{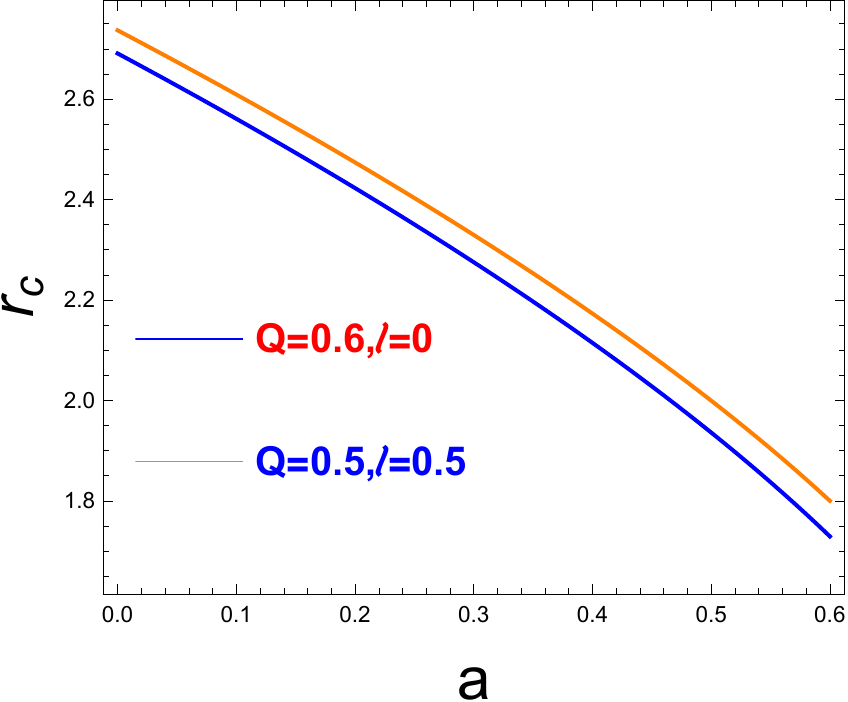}
\endminipage\hfill
			\minipage{0.33\textwidth}
\includegraphics[scale=0.60]{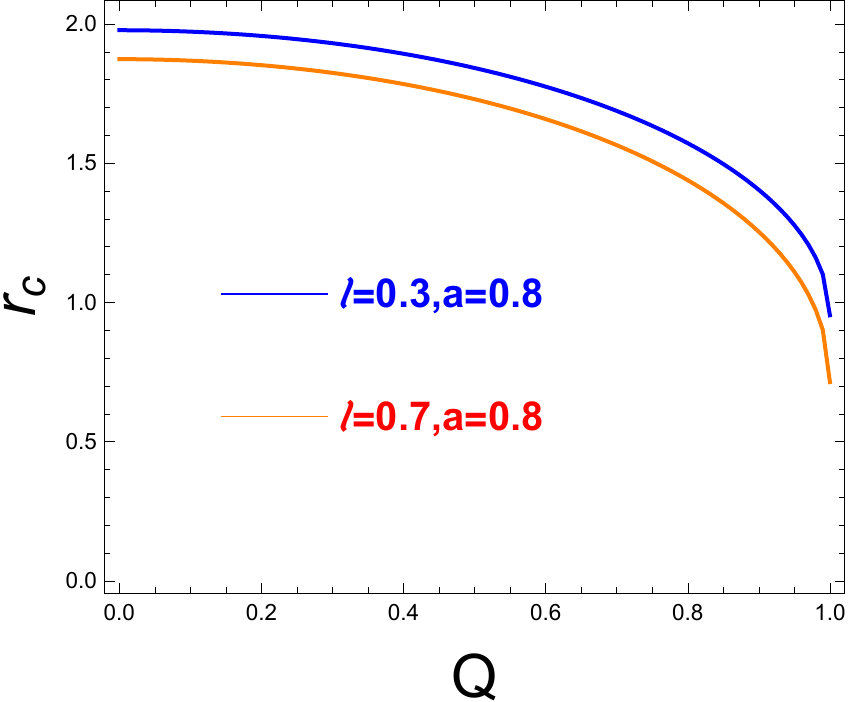}
\endminipage
\caption{Plots showing radii of photon sphere for charged, rotating Kerr-Newman black-bounce metric for different values of black hole parameters. Left most figure displays the variation of photon sphere with respect to $l$ for fixed $a$ and $Q$. The middle figure shows the variation of photon sphere with respect to $a$ for fixed $l$ and $Q$ and the right most figure shows the variation of photon sphere with respect to $Q$ for fixed $a$ and $l$.}
\label{Fig3}
\end{figure}

For Kerr-Newman black-bounce spacetime, we can numerically find out the radius of the photon sphere for fixed values of $l,Q,a$. The variation of photon sphere radius with respect to the different black hole parameters is shown in Figure(\ref{Fig3}). We can clearly see from Fig.(\ref{Fig3}), that for fixed values of $l$ and $a$ as $Q$ increases the radius of photon sphere $r_c$ decreases. Even $r_c$ decreases if we fix any two parameters and increase the other parameter.\\

\begin{figure}[ht!]
\centering
\includegraphics[scale=0.63]{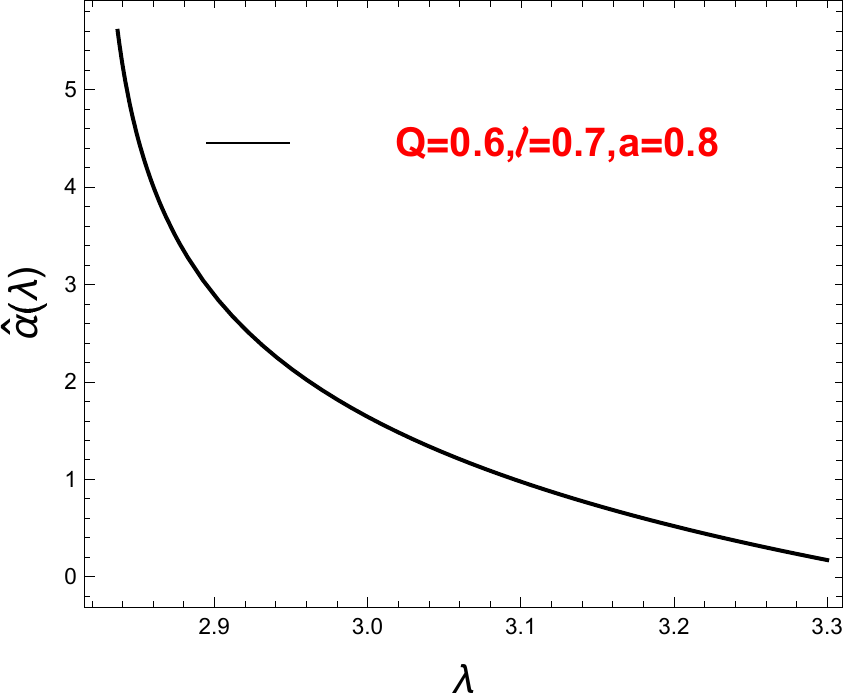}
\includegraphics[scale=0.63]{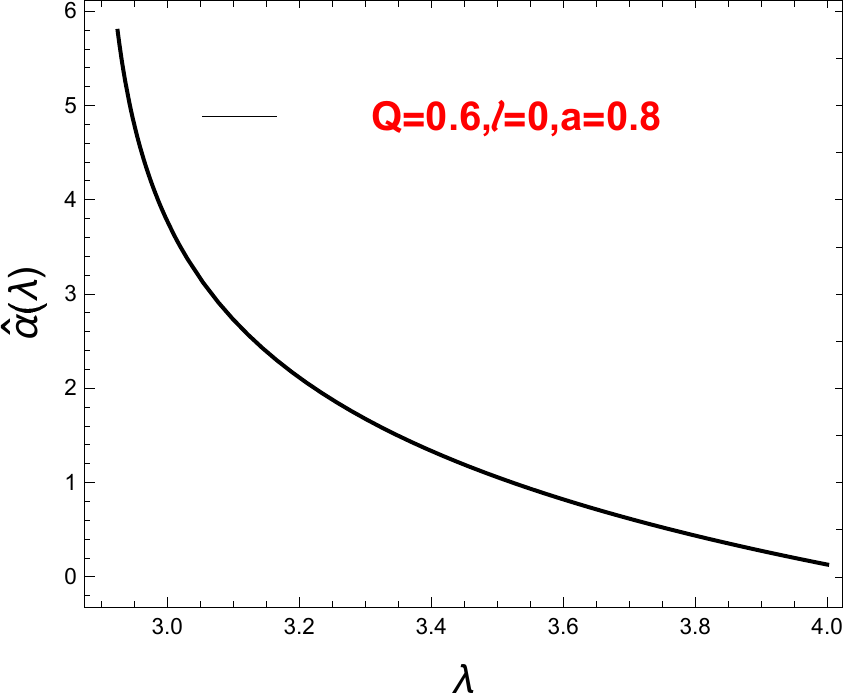}
\includegraphics[scale=0.63]{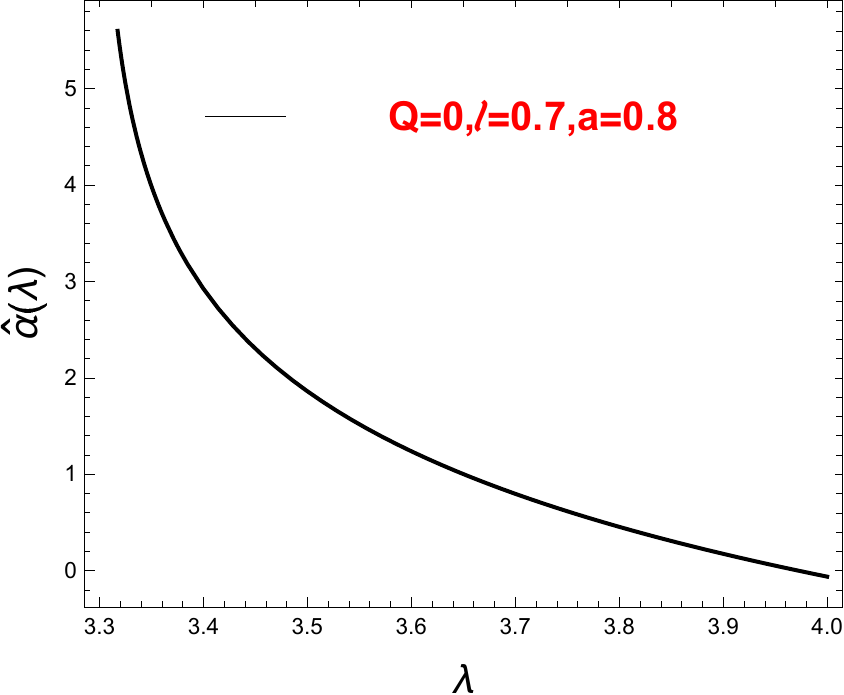}
\caption{Plots showing the variation of the deflection angle $\hat{\alpha}$ with respect to the impact parameter $\lambda$ for fixed values of black hole parameters $Q,l,a$.}
\label{Fig4}
\end{figure}

Armed with these we can now evaluate the divergent integral (\ref{513}),
\begin{eqnarray}
   \phi_{\mathcal{F}}(r_0\approx r_c)&=&\mathcal{R}(z=0,r_c)\int_{0}^1\frac{1}{\sqrt{\gamma_1z+\gamma_2z^2}}\,dz\nonumber\\
   &=&\mathcal{R}(z=0,r_0\approx r_c)\frac{2}{\sqrt{\gamma_2}}\textrm{arcsinh}\Big(\sqrt{\frac{\gamma_2}{\gamma_1}}\Big)\nonumber\\
   &=&\mathcal{R}(z=0,r_0\approx r_c)\frac{2}{\sqrt{\gamma_2}}\log\Big(\frac{\sqrt{\gamma_2}+\sqrt{\gamma_2+\gamma_1}}{\sqrt{\gamma_1}}\Big)\,.\label{521}
\end{eqnarray}
We know that $\phi_{\mathcal{F}}(r_0)$ diverges at $r_0=r_c$ i.e, when the coefficient $\gamma_1=0$. So the idea is to expand the $\gamma_1(r_0)$ around $r_0=r_c$ upto first order and substitute it into (\ref{521}) in order to get the nature of $\phi_{\mathcal{F}}(r_0\approx r_c)$.
\begin{eqnarray}
   \gamma_1(r_0)&=&\frac{\partial \gamma_1}{\partial r_0}\Bigg|_{r_0=r_c}(r_0-r_c)+\mathcal{O}(r_0-r_c)^2\,, \nonumber\\
   \gamma_2(r_0)&=&\gamma_2(r_c)+\frac{\partial \gamma_2}{\partial r_0}\Bigg|_{r_0=r_c}(r_0-r_c)+\mathcal{O}(r_0-r_c)^2\,. \label{522}
\end{eqnarray}
Substituting (\ref{522}) into (\ref{521}) and using the condition (\ref{519}) we get,
\begin{eqnarray}
   \phi_{\mathcal{F}}(r_0\approx r_c)&=&-\tilde{a}\log\Big(\frac{r_0}{r_c}-1\Big)+\tilde{b}+\mathcal{O}(r_0-r_c)\,.\label{523}
\end{eqnarray}
Alternatively, one can write the equation (\ref{523}) in terms of impact parameter as \cite{Bozza:2002af},
\begin{align}
\begin{split} 
  & \hat{\alpha}(\lambda)=-\Bar{a}\log\Big(\frac{\lambda}{\lambda_c}-1\Big)+\Bar{b}+\mathcal{O}(\lambda-\lambda_c) \label{sdeflection},
\end{split}
\end{align}
where the coefficients are,
\begin{eqnarray}
  \Bar{a}&=&\sqrt{\frac{2\mathcal{A}_c\mathcal{B}_c}{\mathcal{A}_c\mathcal{C}_c''-\mathcal{A}_c''\mathcal{C}_c+\lambda_c(\mathcal{A}_c''\mathcal{D}_c-\mathcal{A}_c\mathcal{D}_c'')}}\,,\\
  \Bar{b}&=&-\pi+b_R+\Bar{a}\log\Big(\frac{4\,\gamma_{2c}\,\mathcal{C}_c}{\lambda_c\,\mathcal{A}_c(\mathcal{D}_c+2\lambda_c\mathcal{A}_c)}\Big)\,,\\
  \lambda_c&=&L_c.
\end{eqnarray}
 The variation of the deflection angle (\ref{sdeflection}) with respect to the impact parameter is shown in Fig.(\ref{Fig4}). 
From the Fig.~(\ref{Fig4}) we can see that, for non zero $Q$ and $l$ deflection angle $\hat\alpha(\lambda)$ for a fixed impact parameter decreases. \par
The black hole can be thought of as a lens as gravitational field of which is curving the path of the photons. Let the angular separation between image and lens is ${\theta}=\frac{\lambda}{\boldsymbol{D_{ol}}}$, where $\boldsymbol{D_{ol}}$ is the distance between the lens and the observer. Then the expression for the deflection angle (\ref{sdeflection}) can be written as,
\begin{align}
\begin{split}
  & \hat{\alpha}({\theta})=-\Bar{a}\log(\frac{{\theta}\boldsymbol{D_{ol}}}{\lambda_c}-1)+\Bar{b}+\mathcal{O}(\lambda-\lambda_c)
  \label{426}
\end{split}
\end{align}
where $b_R=\phi_{\mathcal{R}}(r_c)=\int_0^1\mathcal{G}(z,r_c)dz.$ \par 
In order to do the integral we can expand $\mathcal{R}(z,r_c)$  around $ z=0$ and then putting it in (\ref{513}). Formally we get the following expression,
\begin{eqnarray}
   b_R=\frac{1}{\sqrt{\gamma_2}}\int_0^1\Big(\frac{\mathcal{R}(z)}{z}-\frac{\mathcal{R}(0)}{z}\Big)dz\Bigg|_{r_c}=\frac{1}{\sqrt{\gamma_2}}\sum_{n=1}^\infty\frac{1}{n}\frac{\partial^n \mathcal{R}}{\partial z^n}\Bigg|_{z=0}\xrightarrow[]{}\text{ finite}\,.
\end{eqnarray}
When the deflection angle is greater than $2\pi$ we will have multiple images of the source. Using the strong field deflection angle formula given in (\ref{426}) and the lens equation, we can derive the angular radius of the Einstein ring, which is created due to the symmetric lensing of light rays coming from some distant source \cite{PhysRevD.77.124042,Virbhadra:2008ws}. In the next section, we will discuss the observational signatures of deflection angle. 

\section{Observational signatures in strong deflection limit}\label{sec5}
After providing theoretical analysis, in this we will discuss some aspects of observational signatures of deflection angle. The idea is to find the relation between deflection angle and the angular radius of the images (Einstein ring) using lens equation. In this paper we will use the following lens equation given in \cite{Bozza:2001xd},
\begin{eqnarray}
   \beta=\theta-\frac{\boldsymbol{D_{LS}}}{\boldsymbol{D_{OS}}}\Delta\alpha_n\label{61}
\end{eqnarray}
where $\boldsymbol{D_{OS}}=\boldsymbol{D_{OL}}+\boldsymbol{D_{LS}}$, $\beta$ is the angular separation between the source and the lens, and $\boldsymbol{D}_{LS},\boldsymbol{D}_{OS},\boldsymbol{D}_{OL}$ are the distances between  lens to source, observer to source and observer to lens respectively and  $\Delta\alpha_n=\hat{\alpha}(\theta)-2n\pi.$ Now we can write the angular separation between the lens and the  $n^{th}$ image as, $\theta_n=\theta_n^0+\Delta \theta_n$,where $\theta_n^0$ is the angular separation between the image and the lens when the extra deflection angle ($\Delta\theta_n$) over $2n\pi$ is negligible. 
\begin{figure}[b!]
    \centering
 \includegraphics[scale=0.75]{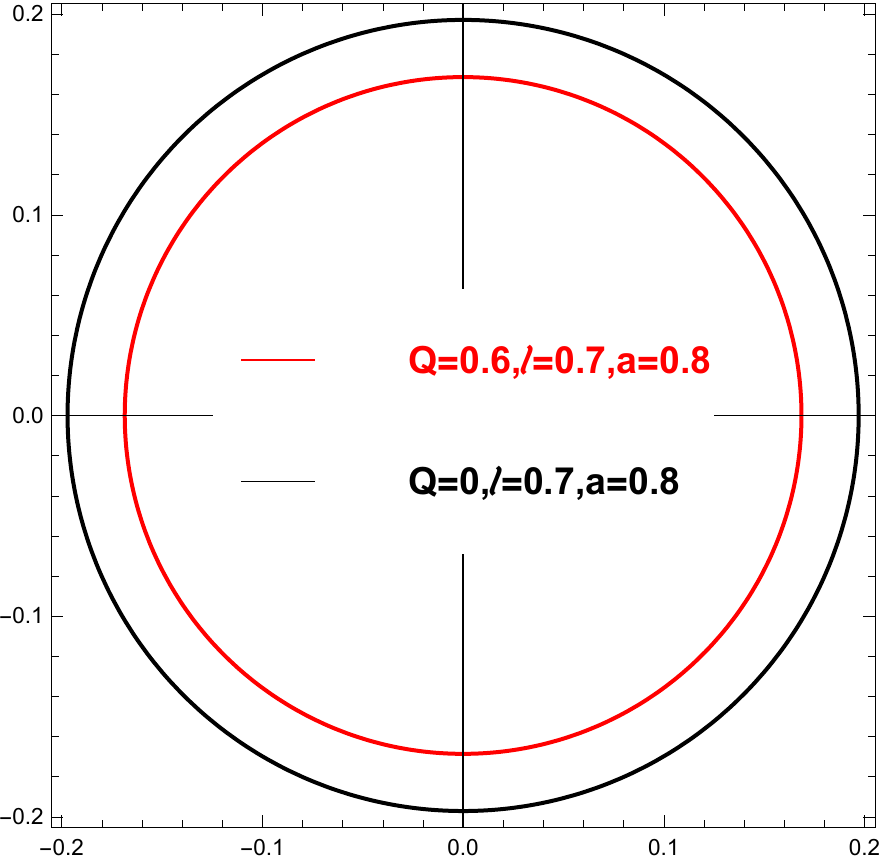}
 \includegraphics[scale=0.75]{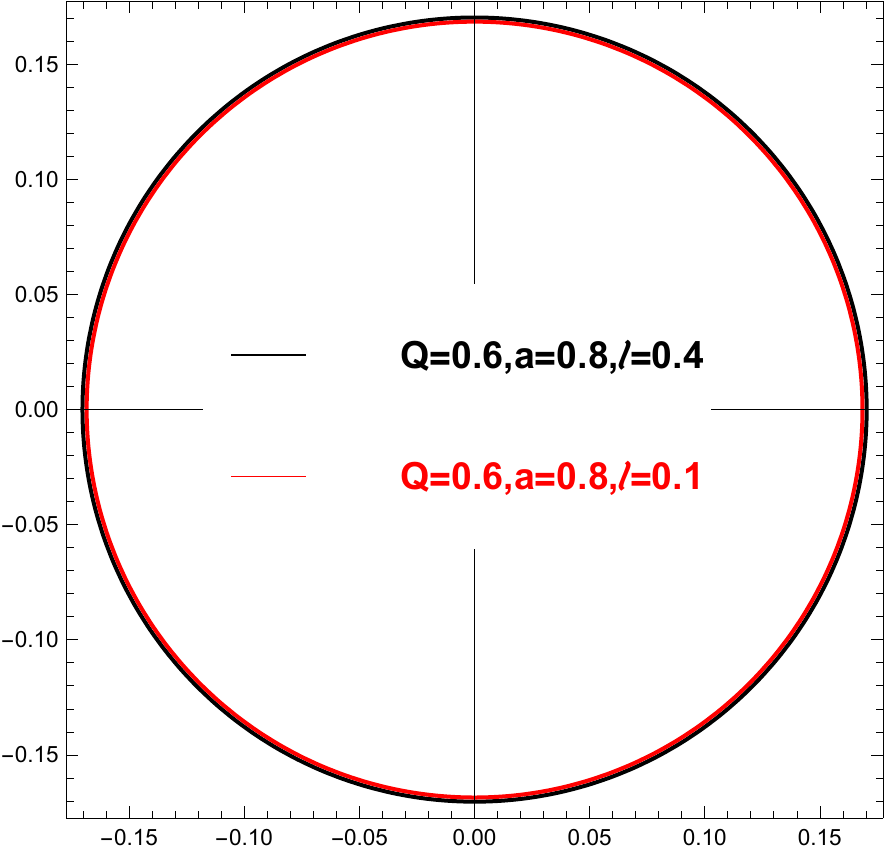}
\caption{Polar plots showing the angular radius $\theta_1$ for different $Q,l,a$ values.}
    \label{Fig5}
    \end{figure}
\par 
For the perfect alignment i.e, when $\beta=0$ and assuming $\Delta\theta_n<<\theta_n^0,$ the angular separation (angular radius) can be written as \cite{Bozza:2002af},
\begin{eqnarray}
   \theta_n^{Einstein}=\frac{\lambda_c}{\boldsymbol{D_{OL}}}\Big[1+\exp\Big(\frac{\Bar{b}}{\Bar{a}}-\frac{2n\pi}{\Bar{a}}\Big)\Big]\,. \label{neq1}
\end{eqnarray}
$n=1$ corresponds to the outermost Einstein ring. We plot some of these rings for different values of $Q,a$ and $l$ in Fig.~(\ref{Fig5}).\par
From the Fig.~(\ref{Fig5}) we can easily see (from the left-most plot in Fig.~(\ref{Fig5}) that \textit{as the charge of the black hole $Q$ increases (for fixed values of $l$ and $a$) the size of the ring decreases. On the other hand, the effect of change of $l$ (for fixed value of $a$ and $Q$) on the ring radius is negligible} as can be easily seen from the right-most plot in Fig.~(\ref{Fig5}). To make this observation more concrete, we further make a comparative study as shown in the Table~~(\ref{Tab1}) and (\ref{Tab2}). We have listed some representative values of the percentage change of the angular radius of the  outermost Einstein rings  with respect to $Q$ (for fixed $a$ and $l$) and $l$ (for fixed $a$ and $Q$) in  Table~(\ref{Tab1}) and (\ref{Tab2}) respectively. These values corroborate perfectly the conclusion drawn above. We can conclude that, in future detections, if one able to resolve angular separation  of various Einstein rings, then that will give us constraint on the charge of the underlying black-bounce metric.  

    \begin{table}[t!]
\centering
\begin{tabular}{|p{1.5cm}|p{0.75cm}|c|c|p{1.5cm}|p{0.75cm}|} 
 \hline
 \multicolumn{2}{|c}{Angular separation}&\multicolumn{1}{|c}{Percentage change}&\multicolumn{1}{|c}{Values of Charge}&\multicolumn{2}{|c|}{Fixed parameters} \\
 \hline
 $\theta_1^{(1)}(Q_2)$&$\theta_1^{(2)}(Q_1)$&$\Delta\theta =\frac{(\theta_1^{(2)}-\theta_1^{(1)})}{\theta_1^{(1)}}\times 100$ & ($Q_1,Q_2$) & $l$ & $a$ \\ 
 \hline
 0.1686 & 0.1970 &16.9013 & (0, 0.6) & 0.7 & 0.8 \\ 
 \hline
0.18548 & 0.21590 & 16.399 & (0.6,0.8) & 0.4 & 0.5 \\
 \hline
 0.215903 & 0.232365 & 7.62463 & (0.4,0.6) & 0.4 & 0.5\\
 \hline
\end{tabular}
 \caption{Percentage change of the angular radius of first Einstein ring for different values of charge $Q$ for fixed $a$ and $l$.}
\label{Tab1}
\end{table}
\begin{table}[t!]
\centering
\begin{tabular}{|p{1.5cm}|p{0.75cm}|c|c|} 
 \hline
 \multicolumn{2}{|c}{Angular separation}&\multicolumn{1}{|c}{Percentage change}&\multicolumn{1}{|c|}{\makecell{Values of\\ regularization\\ parameter}} \\
 \hline
 $\theta_1^{(1)}(l_1)$&$\theta_1^{(2)}(l_2)$&$\Delta\theta =\frac{(\theta_1^{(2)}-\theta_1^{(1)})}{\theta_1^{(1)}}\times 100$ & $(l_1,l_2)$ \\ 
 \hline
 0.16856 & 0.1703 & 1.07059 & (0.2,0.4)  \\ 
 \hline
 0.16856 & 0.16860 & 0.0186897 & (0.2,0.7) \\
 \hline
 0.168565 & 0.168568 & 0.001963 & (0.1,0.2)\\
 \hline
\end{tabular}
 \caption{Percentage change of the angular radius of first Einstein ring for different values of regularisation parameter $l$ for  $a=0.8$ and $Q=0.6$.}
\label{Tab2}
\end{table}
\par
%Till now, we have considered the equatorial lensing for perfect alignment, i.e $\beta=0$ as mentioned above (\ref{neq1}). However, one can generalize this result for $\beta \neq 0.$ Then the angular separation of the Einstein ring can be written as \cite{Bozza:2002af},
%\begin{equation}
% \theta_{n,\beta}^{Einstein}=\frac{\lambda_c}{\boldsymbol{D_{OL}}}\Big[1+\exp\Big(\frac{\Bar{b}}{\Bar{a}}+\frac{\beta}{\Bar{a}}-\frac{2n\pi}{\Bar{a}}\Big)\Big]\,.
%\end{equation}
%In this section, we first consider the equatorial lensing for different $\beta$. Below we quote the angular separation of the Einstein ring for some representative values of $\beta.$\par 
%Case 1: $\beta=\pi$, i.e., source and observer lie on the same side of the lens. Then,
%\begin{eqnarray}
 %\theta_{n,\pi}^{Einstein}=\frac{\lambda_c}{\boldsymbol{D_{OL}}}\Big[1+\exp\Big(\frac{\Bar{b}}{\Bar{a}}+\frac{\pi}{\Bar{a}}-\frac{2n\pi}{\Bar{a}}\Big)\Big]\,.
%\end{eqnarray}
\par
\newpage
\section {Strong deflection analysis of non-equatorial lensing}\label{Sec6}
In Section~\ref{sec5}, we have investigated the lensing in the equatorial plane. In this section, we will further extend our study for the non-equatorial lensing at a small inclination to get a complete picture, as there is no spherical symmetry for the metric under consideration. We will follow \cite{Bozza:2002af} closely to carry out the analysis. In case of equatorial lensing we only need the one-dimensional lens equation. However, to investigate the caustic structure and magnification for non equatorial case, we need a two-dimensional lens equation given in \cite{Bozza:2002af}. We assume that inclination is  $\theta=\frac{\pi}{2}-\psi$, where $\psi$ is very small. 
%We will mainly use \cite{Bozza:2002af} to calculate the deflection angle at small polar inclination. 
\par 
For non-equatorial plane we can not set the carter constant $\mathcal{K}$ to 0 i.e, $\mathcal{K}\neq 0$. Rather for the small inward inclination angle we can write down the constants in terms of inclination angle $\psi$ as follows,
\begin{eqnarray}
 L&\approx& \lambda\\
 \mathcal{K}&\approx&h^2+(\lambda^2-a^2) \psi_0^2\,\,, \text{with}\,\, \psi_0\approx \frac{h}{\lambda}
\end{eqnarray}
Now using the $\theta$ and $\phi$ geodesic equation in \ref{211} and requiring $\psi$ to be small we will have,
\begin{align}
    \begin{split}
        \frac{d\psi}{d\phi}= \omega(r(\phi))\sqrt{\hat{\psi}^2-\psi^2} \,\,,\text{with}\,\,\hat{\psi}=\sqrt{\frac{h^2}{\lambda^2-a^2}+\psi_0^2}
    \end{split}
\end{align}
We are interested in the quantity (deflection angle),
\begin{eqnarray}
 \bar{\phi_f}=\int_0^{\phi_f}d\phi\,\, \omega(\phi)
\end{eqnarray}
where $\phi_f$ is the total azimuthal shift. Then the deflection angle can be written as
\begin{eqnarray}
 \bar{\phi_{f}}=2\,\int_{r_0}^\infty dr \, \omega(r) \frac{d\phi}{dr}=\int_0^1\,dz\,\,\omega(r(z)) \mathcal{R}(z,r_0) \mathcal{F}(z,r_0)
\end{eqnarray}
where,
\begin{eqnarray}
 \omega(r)= \bar{\lambda}\,\frac{a^2+\sqrt{r^2+l^2}(\sqrt{r^2+l^2}-2)}{\sqrt{r^2+l^2}\Big(2\,a+\lambda(\sqrt{r^2+l^2}-2)\Big)+Q^2(\lambda-a)}
\end{eqnarray}
with, $\bar{\lambda}=\sqrt{\lambda^2-a^2}$.
We can follow the same method to get the divergent nature of the deflection angle as the function $\omega(r)$ is free of singularities and is given by,
\begin{eqnarray}
 \bar{\phi_{f}}=-\hat{a}\log\Big(\frac{\lambda}{\lambda_c}-1\Big)+\hat{b}.
\end{eqnarray}
Here,
\begin{eqnarray}
 \hat{a}&=&\frac{\omega(z=0, r_c)\mathcal{R}(z=0,r_c)}{2\sqrt{\gamma_{2c}}},\\
 \hat{b}&=&-\pi+\hat{b}_R+\hat{a}\log\Big(\frac{4\,\gamma_{2c}\,\mathcal{C}_c}{\lambda_c\,\mathcal{A}_c(\mathcal{D}_c+2\lambda_c\mathcal{A}_c)}\Big)\,,
 \end{eqnarray}
where,
\begin{eqnarray}
 \hat{b}_R=\int_0^1\,dz\,[\omega(z,r_c)\mathcal{R}(z,r_c)\mathcal{F}(z,r_c)-\omega(z=0,r_c)\mathcal{R}(z=0,r_c)\mathcal{F}(z,r_c)].
\end{eqnarray}\par
\begin{figure}[htb!]
    \centering
 \includegraphics[scale=0.50]{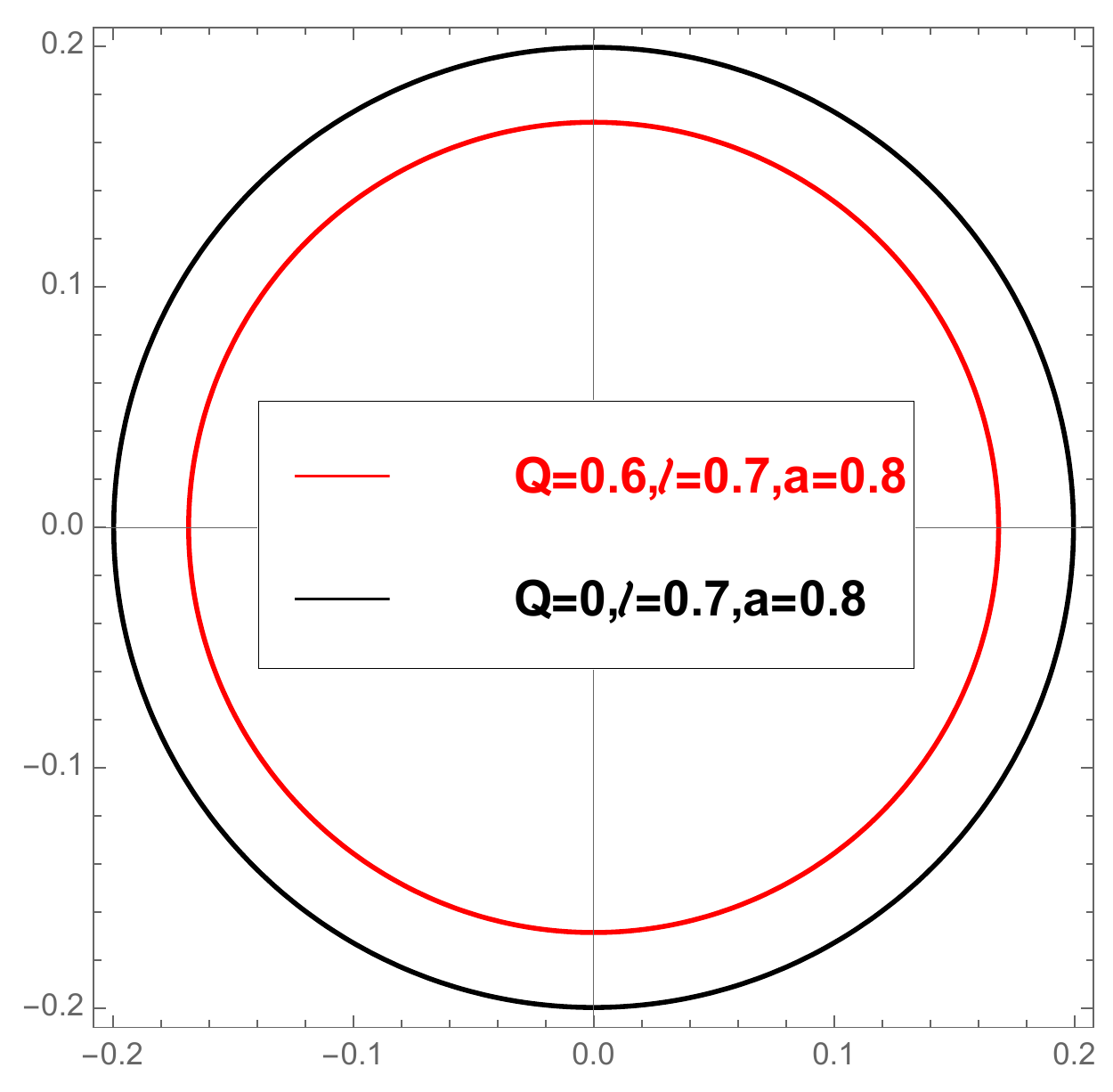}
 \includegraphics[scale=0.50]{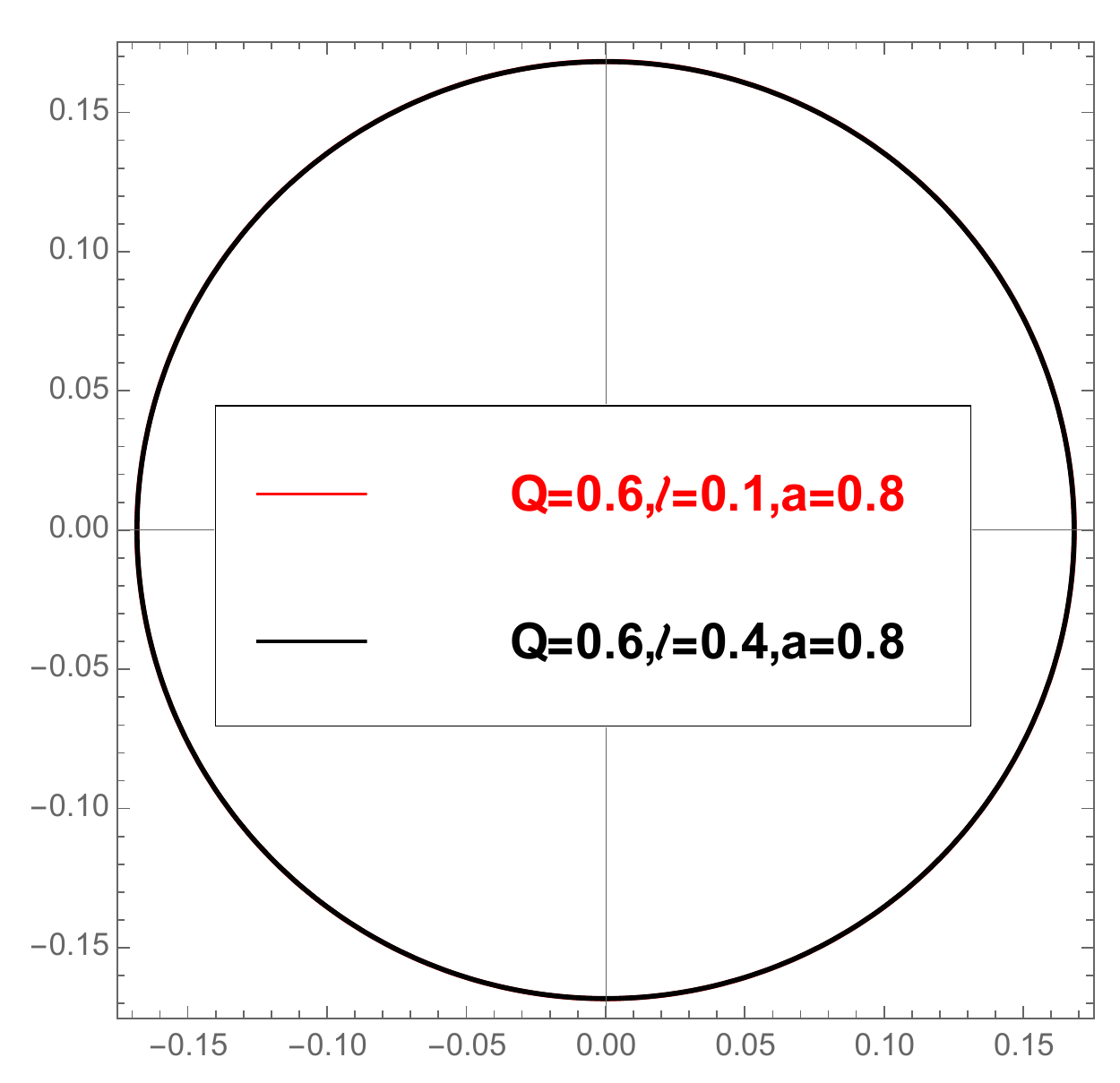}
\caption{Polar plots showing the angular radius $\theta_1$ for different $Q,l,a$ values.}
    \label{Fig6}
    \end{figure}
\textit{Fig~(\ref{Fig6}) shows that the angular radius of first Einstein ring  depends on the charge parameter of the black bounce metric considerably but the dependence on the regularisation parameter is minimal as we see in case of equatorial lensing.}\par
Our interest is to find the position of the caustics, where the magnification diverges, and is given by \cite{Bozza:2002af},
$\bar{\gamma_k}=-\bar{b}+\bar{a}(\hat{b}-k\,\pi)$. Here $k$ is a positive integer. For each values of $k,$ we have one caustic for the direct photon and one for the retrograde photon. For $k=1$, we will find the caustic point in the weak field limit and for $k\ge 1$ corresponds to the strong field limit, and we are interested in this regime.
 \begin{table}[]
\centering
\begin{tabular}{|p{2cm}|p{0.75cm}|c|c|p{1.5cm}|} 
 
 \hline
 ($l,a$)&$Q$&$ \bar{\gamma_2}$ \\ 
 \hline
 (0.4,0.5) & 0.5 & -6.62\\
 \hline
  (0.4,0.5) & 0.6 & -5.10\\
  \hline 
  (0.4,0.5) & 0.7 & -0.16\\
  \hline
   (0.4,-0.5) & 0.5 & -5.69\\
   \hline
   (0.4,-0.5) & 0.6 & -5.84\\
   \hline
   (0.4,-0.5) & 0.7 & -5.90\\
   \hline
\end{tabular}
 \caption{Angular position of the second caustic point for different charge parameter $Q$. The first three entities are for direct photon and the last three are for retrograde photon.}
\label{Tab5}
\end{table}
In Table ~(\ref{Tab5}), we investigate the variation of the second caustic point with respect to $Q$ for fixed $l$ and $a$.
\begin{table}[]
\centering
\begin{tabular}{|p{2cm}|p{0.75cm}|c|c|p{1.5cm}|} 
 
 \hline
 ($Q,a$)&$l$&$ \bar{\gamma_2}$ \\ 
 \hline
 (0.5,0.5) & 0.4 & -6.62\\
 \hline
  (0.5,0.5) & 0.5 & -6.68\\
  \hline 
  (0.5,0.5) & 0.6 & -6.72\\
  \hline
   (0.5,-0.5) & 0.7 & -5.92\\
   \hline
   (0.5,-0.5) & 0.8 & -5.98\\
   \hline
   (0.5,-0.5) & 1.0 & -6.20\\
   \hline
\end{tabular}
 \caption{Angular position of the second caustic point for different regularisation parameter $l$. The first three entities are for direct photons, and the last three are for retrograde photons.}
\label{Tab6}
\end{table}
In Table ~(\ref{Tab6}), we investigate the variation of the second caustic point with respect to $l$ for fixed $Q$ and $a$.
%\newpage
Comparing Table ~(\ref{Tab5}) and (\ref{Tab6}) we can conclude that the change of caustic points for different $Q$ and different $l$ is not that robust.

\section{Analysis of the observables for the shadow}\label{sec6}
To complement our discussions in the previous section, we also investigate the size of shadow cast by the underlying spacetime and various associated observable quantities and their dependence on the charge of the black-bounce metric. In light of recent EHT observations, particularly as it is expected that the presence of charge affects the size of the shadow contour significantly \cite{EventHorizonTelescope:2021dqv}.  \par
For this, we first need to find out the region of unstable spherical light rays that acts as a boundary beyond which the photons will be captured by the black hole. This then leads to a black patch as no light rays reach the observer’s line of sight and serve as the boundary of the black patch, which is the so-called black hole shadow. Sometimes they are also called as
photon spheres. For rotating black holes there are two circular photon orbits and the region between them is called the photon region. The photon region can be found by solving $\dot{r}=\ddot{r}=0$. From equation (\ref{211}) we get,
\begin{eqnarray}
0&=& \mathcal{R}(r):=\frac{R(r)}{E^2}=[(r^2+a^2+l^2)-a\, \lambda]^2-\Delta(r)[\eta+(\lambda-a)^2] \,\,\,\text{and},\label{71}\\
0&=&\mathcal{R}'(r)=4\,r[(r^2+a^2+l^2)-a\,\lambda]-\Delta'(r)[\eta+(\lambda-a)^2].\label{72}
\end{eqnarray}
where, $\lambda=\frac{L}{E}$ and $\eta=\frac{\mathcal{K}}{E^2}.$ Now solving (\ref{71}) and (\ref{72}) we find the two constants $\lambda$ and $\eta$ and are given by,
\begin{align}
    \begin{split}
        & \lambda=\frac{(a^2 +l^2+r^2)\Delta'(r)-4 r \Delta(r)}{a \Delta'(r)}\Big|_{r=r_p}\label{73}
    \end{split}
\end{align}
and
\begin{align}
    \begin{split}
   \displaystyle{\eta=\frac{16 a^2 r^2 \Delta(r)-l^4\Delta'(r)^2-2 l^2 r^2 \Delta'(r)^2+8 l^2\, r \Delta(r) \Delta'(r)-r^4 \Delta'(r)^2+8 r^3 \Delta(r) \Delta'(r)-16 r^2 \Delta(r)^2}{a^2 \Delta'(r)^2}
    \Big|_{r=r_p}}.\label{74}
    \end{split}
\end{align}
 Now substituting (\ref{73}) and (\ref{74}) in the positivity condition of the $\theta$ equation of motion (\ref{211}) we will get the photon region,
\begin{eqnarray}
 \tilde{\mathcal{H}}:\eta+\cos^2\theta\Big(a^2-\frac{\lambda^2}{\sin^2 \theta}\Big)\Big|_{r=r_p}\ge 0
\end{eqnarray}
$\tilde{\mathcal{H}}$ denotes the photon region and $r_p$ denotes the values of radial coordinate $r$ on the shadow contour.\par
In order to investigate the shadow structure we can define two asymptotic celestial coordinate at a given inclination angle $\theta_0$ as \cite{Wei:2019pjf},
\begin{eqnarray}
\alpha&=&-\lambda \csc(\theta_0)\nonumber\\
\beta&=&\pm \sqrt{\eta+a^2 \cos(\theta_0)^2-\lambda^2 \cot(\theta_0)^2}
\end{eqnarray}
For simplicity we take $\theta_0=\frac{\pi}{2}$. Then the coordinates become,
\begin{align}
\begin{split}
 \alpha(r_p)=-\lambda(r_p),\,\,
 \beta(r_p)=\pm \sqrt{\eta(r_p)}
 \end{split}
\end{align}
\begin{figure}[t!]
    \centering
 \includegraphics[scale=0.50]{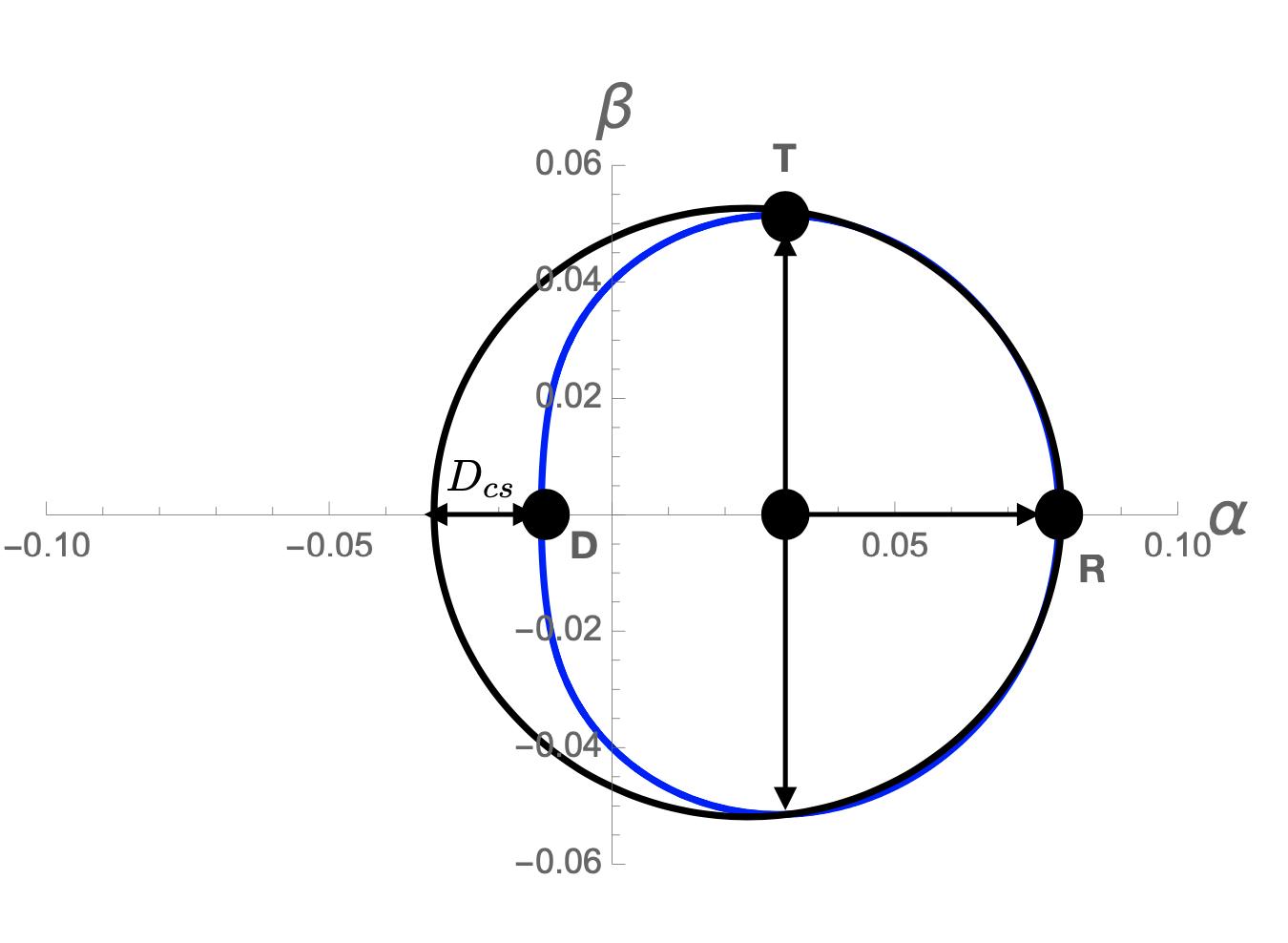}
\caption{Figure displays various characteristic points on a typical shadow contour}
    \label{Fig7}
    \end{figure}
Next, to analyze the effect of charge on the topology of the shadow contour, we compute the observable quantities \cite{Wei:2019pjf}. Particularly we focus on the radius of the shadow and the distortion parameter. In order to investigate these, we first consider three characteristic points of the shadow contour as shown in Fig.~(\ref{Fig7}). We follow the analysis of \cite{Wei:2019pjf}. In order to find the Top (`{\bf T}') point, we need to find the root of, 
\begin{eqnarray}
 (\partial_{r_{p}} \beta)\Big|_{\theta_0=\frac{\pi}{2}}=0 \implies (\partial_{r_{p}} \eta)\Big|_{\theta_0=\frac{\pi}{2}}=0\,. 
 \label{68}
\end{eqnarray}
At ${\bf T},$ $\beta$ takes the maximum value. The condition (\ref{68}) gives,
\begin{align}
    \begin{split}
      &  \Big(a^2 \Big(\sqrt{l^2+r_p^2} \left(2 Q^2+r_p^2\right)+l^2 \Big(\sqrt{l^2+r_p^2}\\ &
      -3\Big)-3 r_p^2\Big)-l^6+2\, Q^4 \sqrt{l^2+r_p^2}+l^2 \Big(Q^2 \left(7 \sqrt{l^2+r_p^2}-4 r_p^2-9\right)\\ & 
      +9 \sqrt{l^2+r_p^2}-3\, r_p^2 \left(-4 \sqrt{l^2+r_p^2}+r_p^2+8\right)\Big)
      +7 Q^2 r_p^2 \sqrt{l^2+r_p^2}+9\, r_p^2 \sqrt{l^2+r_p^2}+6 \, r_p^4 \sqrt{l^2+r_p^2}\\ &+l^4 \left(6 \sqrt{l^2+r_p^2}-2 \left(Q^2+6\right)-3 \, r_p^2\right)-2 \left(Q^2+6\right) r_p^4-9 Q^2 r_p^2-r_p^6\Big)=0\,. \label{699}
    \end{split}
\end{align}
\begin{table}[t!]
\centering
\begin{tabular}{|c|c|c|c|} 
 \hline
 $Q$ & $R_s$ & $\delta_s=\frac{D_{cs}}{R_s}$ & $(\alpha_R-\alpha_D)$ \\
 \hline
 0 & 5.196 & 0.03038  & 10.2344 \\ 
 \hline
 0.1 & 5.1874 & 0.03072 & 10.2156\\ 
 \hline
 0.2 & 5.16124 & 0.03177 & 10.1585\\
 \hline
 0.3 & 5.11679 & 0.03369 & 10.0612\\
 \hline
 0.4 & 5.05298 & 0.0367677 & 9.92017\\
 \hline
 0.5 & 4.96791  & 0.0415804 & 9.72926\\
 \hline
 0.6 & 4.85869 & 0.0493605 & 9.72657\\
 \hline
 0.7 & 4.72068 & 0.0631616 & 9.1432 \\
 \hline
 0.8 & 4.54599 & 0.0939478 & 8.66489\\
 \hline
  \end{tabular}
 \caption{The above table shows how the shadow distortion parameter changes with charge $Q$ for $a=0.5$ and $l=0.6.$}
\label{Tab3}
\end{table}
We denote the solution of $r$ corresponding to the point ${\bf T}$ coming from (\ref{699}) by $\tilde r_p.$ \par 
To find the coordinates of other two points ${\bf D}$ and ${\bf R}$ we need to set $\beta=0$ as evident  from Fig.~(\ref{Fig6}). The two real positive roots ($r_p^{+},\,\, r_p^{-}$) of it  characterize these two points ${\bf D}$ and ${\bf R}$ are respectively. So we have the following, 
\begin{eqnarray}
 T:&=&\Big(\alpha_{T}(\tilde{r_p}),\beta_{T}(\tilde{r_p})\Big)\nonumber\\
 R:&=&\Big(\alpha(r_p^+),0\Big)\\
  D:&=&\Big(\alpha(r_p^-),0\Big)
 \end{eqnarray}
 For more details interested readers are referred \cite{Hioki:2009na,Gralla:2019xty,Wei:2019pjf,Chowdhuri:2020ipb}. Further we compute two physical quantities namely \cite{Hioki:2009na,Gralla:2019xty,Wei:2019pjf,Chowdhuri:2020ipb}, 
 \begin{align}
    \textrm{Radius\,\, of\,\, the\,\, shadow: }\,\, R_s = \Delta \beta
 \end{align}
 where $\Delta\beta=\beta (\tilde{r_p})$
 and 
 \begin{align}
    \textrm{Distortion\,\, parameter: }\,\, \delta_s =\frac{D_{cs}}{R_s},\quad  D_{cs}=2\, R_s-(\alpha_R-\alpha_D).
 \end{align}
$R_s$ is basically the radius of the reference circle as shown by the black curve in Fig~(\ref{Fig6}) and $D_{cs}$ measures the deviation from circularity when the shadow gets distorted whenever there are non-vanishing $a.$ A typical distorted shadow contour is shown by blue curve in Fig.~(\ref{Fig6}). \par We can use these two quantities $(R_s,\,\delta_s)$ we can make contact with the observational data. We list values of these quantities in Table~(\ref{Tab3}). As discussed before, we are particularly interested in investigating the effect of charge on the size of the shadow contour. \textit{It is evident from the data listed in Table~(\ref{Tab3}) that, for fixed values of $a$ and $l$ as we increase the value of the charge $Q$ the shadow becomes more distorted. From this analysis we can clearly see the effect of charge on the observational appearance of  this black-bounce metric. \footnote{In principle, we can calculate the distortion parameter different inclination angles $\theta_0,$, but that will not change the conclusion drawn in this section.}} But the change in the distortion parameter $\delta_s$ as evident from the Table~(\ref{Tab3}) is such that the EHT can not resolve this with its current resolution power \cite{EventHorizonTelescope:2019dse}. But in the future, we hope that with the increase of resolution power, this change can be detected and help provide tighter constraints on the underlying parameter space. 
 
\section{Discussions}\label{sec7}
In summary, the motion of light rays around Kerr-Newman black-bounce spacetime has been studied in this paper. We first study the null geodesics and the innermost circular orbit in the bounce spacetime. From this investigation,  we found that the black hole parameter $l$ gives an extra attractive effect to the photon motion. Effectively the position of photon sphere has been pulled back in comparison with the Kerr-Newman black hole along with the repulsive effect caused by the charge of the black hole, which has a direct consequence on the observational appearance of the black hole. \textit{We then give an analytical description of the equatorial deflection angle perturbatively in $l$, in terms of different kinds of elliptic integrals.} The non-zero $l$ decreases the deflection angle for a fixed impact parameter. Our work gives a general methodology to study deflection angle analytically.  \par 
Throughout the paper, we give a comparative study of different observational signatures in Kerr-Newman black-bounce spacetime as well as a detailed analytical study. In this context, we first study the strong deflection analysis of the equatorial deflection angle, and this has direct observational consequences as this gives information about the Einstein rings. We conclude this analysis by investigating the first-order Einstein ring using the lens equation. From our analysis, we conclude that the size of the Einstein ring prominently depends on the black hole charge.\textit{ We observed that, as we decrease the charge ($Q $), the size of the ring increases. On the other hand, the effect of the regularisation parameter ($l$) on the ring radius is minimal compared to the effect of charge for a fixed value of spin parameter $a$. Furthermore, we also investigate the Einstein ring in case of small polar inclination. We observed that the dependence on charge ($Q$) and regularisation parameter ($l$) remain the same as in the case of equatorial lensing.} Also, we showed that the change in the position of caustic points with the change of charge ($Q$) and regularisation parameter ($l$) is not that considerable. One can apply this to different black hole spacetime to get an exact analytical expression.
\par
Furthermore, motivated by the recent observations from EHT \cite{EventHorizonTelescope:2021dqv}, we compute the distortion parameter for the shadow cast by this metric and investigate its dependence on the charge parameter $Q$ to complement our analysis for the Einstein ring. We hope that these results will help us to effectively constrain the parameter space of this black-bounce metric in the near future with the increase of the resolving power of the telescope. Note that we have chosen the parameter space such that the horizon exists but it would be interesting to investigate the shadow structure when we do not have horizon and $r_{\rm{p}}=r_{\rm{throat}}.$  In that case, there might be a possibility of getting a significant change in the distortion parameter quantifying the topology of the shadow due to the regularization parameter $l$ \cite{Kasuya:2021cpk}.\footnote{We thank Naoki Tsukamoto for pointing out this reference.}
\par
Several interesting avenues can be perused in future. Our work can be extended for other black-bounce spacetime e.g \cite{Barrientos:2022avi}. One can study multilevel images in this spacetime \cite{Hou:2022gge}. First of all, our analytical study presented in Sec.~\ref{sec3} requires a small $l$ expansion and also, the study of non-equatorial lensing in Sec.~\ref{Sec6} assumes a small inclination angle. It will be interesting to go beyond this regime, which will require a completely numerical study. We leave this investigation for future study. One can also study multilevel images in this spacetime \cite{Hou:2022gge}. From that analysis,  From that analysis, one can predict how much precision of resolution is needed for distinguishing different Einstein rings. Apart from that, one can investigate strong gravitational lensing in an environment of thin accretion, or plasma \cite{Bisnovatyi-Kogan:2017kii}. Last but not least, another important quantity from the image analysis point of view that can be analysed is the two-point correlation function of intensity fluctuations on the photon ring, which arises due to the photon moving around multiple orbits around the central object. We hope to make a similar analysis for this black-bounce metric along the line of \cite{Hadar:2020fda}. These studies will further help us to make contact with realistic astrophysical scenarios. We hope to report about some of these issues in the near future. \newpage

\section*{Acknowledgements}

A.B is supported by Start-Up Research Grant (SRG/2020/001380), Mathematical Research Impact Centric Support Grant (MTR/2021/000490) by the Department of Science and Technology Science and Engineering Research Board (India) and Relevant Research Project grant (202011BRE03RP06633-BRNS) by the Board Of Research In Nuclear Sciences (BRNS), Department of atomic Energy, India. Authors thank Abhishek Chowdhuri and Mostafizur Rahman for useful discussions.

\appendix
 \section{Appendix 1: Details of the integrals used in Section~(\ref{sec3})}\label{App}
 In this section we will show how to arrive the elliptic closed form of the integrals that is used in Section~\ref{sec3}).  We have encountered mainly two types of integral given by,
 \begin{eqnarray} 
 I_1=\int_0^{u_2}du \frac{u^2}{(u-u_+)\sqrt{-(u-u_1)(u-u_2)(u-u_3)(u-u_4)}}\label{A1}
 \end{eqnarray}
and
\begin{eqnarray}
 I_2:=\int_0^{u_2}du \frac{u^3}{(u-u_+)\sqrt{-(u-u_1)(u-u_2)(u-u_3)(u-u_4)}}\label{A2}
\end{eqnarray}
with $u_4>u_3>u_2>0>u_1$.\par
We can write $I_1$ (defined in (\ref{A1}) as,
\begin{align}
    \begin{split}
        &  I_1=\int_0^{u_2}du\frac{1}{\sqrt{-(u-u_1)(u-u_2)(u-u_3)(u-u_4)}}\Bigg[u_++u+\frac{u_+^2}{u-u_+}\Bigg]\\ &
        =\boldsymbol{N_1+N_2+N_3}\,,
    \end{split}
\end{align}
where
\begin{align}
    \begin{split}
     \boldsymbol{N_1}=\int_0^{u_2}du\frac{u_+}{\sqrt{-(u-u_1)(u-u_2)(u-u_3)(u-u_4)}}
    =u_+g\Big(F\big(\frac{\pi}{2},k\big)-F(\phi,k)\Big),  \end{split}
\end{align}
\begin{align}
    \begin{split}
    \boldsymbol{N_2}& =\int_0^{u_2}\, du\frac{u}{\sqrt{-(u-u_1)(u-u_2)(u-u_3)(u-u_4)}}\\ &
    =\int_0^{u_1}\, du\frac{u}{\sqrt{-(u-u_1)(u-u_2)(u-u_3)(u-u_4)}}+\int_{u_1}^{u_2}du\frac{u}{\sqrt{-(u-u_1)(u-u_2)(u-u_3)(u-u_4)}}\\ &
    =-g\, u_1\int_0^{t_1}\, dt\frac{1-\frac{u_4\alpha^2}{u_1}sn^2t}{1-\alpha^2t}+g\,u_1\int_0^{s_1}dt\frac{1-\frac{u_4\alpha^2}{u_1}sn^2t}{1-\alpha^2t}\\ &
    =g\,u_1\Big[\Big(1-\frac{u_4}{u_1}\Big)\Big(\Pi(\frac{\pi}{2},\alpha^2,k)-\Pi(\phi,\alpha^2,k)\Big)+\frac{u_4}{u_1}\Big(F(\frac{\pi}{2},k)-F(\phi,k)\Big)\Big]\\ &
    =g\Big[\Big(u_1-u_4\Big)\Big(\Pi(\frac{\pi}{2},\alpha^2,k)-\Pi(\phi,\alpha^2,k)\Big)+u_4\Big(F(\frac{\pi}{2},k)-F(\phi,k)\Big)\Big]
    \end{split}
    \end{align}
    and
\begin{align}
    \begin{split}
   \boldsymbol{N_3}&=\int_0^{u_2}du\frac{u_+^2}{(u-u_+)\sqrt{-(u-u_1)(u-u_2)(u-u_3)(u-u_4)}}\\ &
    =u_+^2\frac{g}{(u_+-u_1)}\Big[\frac{u_1-u_4}{u_+-u_4}\Big(\Pi(\phi,\alpha_3^2,k)-\Pi(\frac{\pi}{2},\alpha_3^2,k)\Big)+\frac{u_+-u_1}{u_+-u_4}\Big(F(\phi,k)-F(\frac{\pi}{2},k)\Big)\Big]\,.
     \end{split}
\end{align}
Finally we get,
\begin{align}
    \begin{split}
     &   I_1=\int_0^{u_2}du \frac{u^2}{(u-u_+)\sqrt{-(u-u_1)(u-u_2)(u-u_3)(u-u_4)}}\\ & =u_+g\Big(F\Big(\frac{\pi}{2},k\Big)-F(\phi,k)\Big)+g\Big[\Big(u_1-u_4\Big)\Big(\Pi(\frac{\pi}{2},\alpha^2,k)-\Pi(\phi,\alpha^2,k)\Big)+u_4\Big(F(\frac{\pi}{2},k)-F(\phi,k)\Big)\Big]\\ &
     +u_+^2\frac{g}{(u_+-u_1)}\Big[\frac{u_1-u_4}{u_+-u_4}\Big(\Pi(\phi,\alpha_3^2,k)-\Pi(\frac{\pi}{2},\alpha_3^2,k)\Big)+\frac{u_+-u_1}{u_+-u_4}\Big(F(\phi,k)-F(\frac{\pi}{2},k)\Big)\Big]\,.\label{87}
    \end{split}
\end{align}
Next we focus on $I_2$ defined in (\ref{A2}).
\begin{align}
\begin{split}
 & I_2= \int_0^{u_2}du \frac{u^3}{(u-u_+)\sqrt{-(u-u_1)(u-u_2)(u-u_3)(u-u_4)}} \\ &
 =\int_0^{u_2}du\frac{1}{\sqrt{-(u-u_1)(u-u_2)(u-u_3)(u-u_4)}}\Big[(u-u_+)^2+3u_+u+\frac{u_+^3}{u-u_+}\Big]\\ &
 =\boldsymbol{M_1+M_2+M_3}.
\end{split}
\end{align}
Now we will again do the integrals piece-wise and get,
\begin{align}
\begin{split}
 \boldsymbol{M_1} & = \int_0^{u_2}du\frac{u^2}{\sqrt{-(u-u_1)(u-u_2)(u-u_3)(u-u_4)}}-2u_+\int_0^{u_2}du \frac{u}{\sqrt{-(u-u_1)(u-u_2)(u-u_3)(u-u_4)}}\\ &
+\int_0^{u_2}\frac{du}{\sqrt{-(u-u_1)(u-u_2)(u-u_3)(u-u_4)}}\\ &
= g\Big[u_4^2\Big(F(\frac{\pi}{2},k)-F(\phi,k)\Big)+2u_4(u_1-u_4)\boldsymbol V_1+(u_1-u_4)^2\boldsymbol{V_2}\Big]\\ &-
2 \, u_+g\Big[\Big(u_1-u_4\Big)\Big(\Pi(\frac{\pi}{2},\alpha^2,k)-\Pi(\phi,\alpha^2,k)\Big)+u_4\Big(F(\frac{\pi}{2},k)-F(\phi,k)\Big)\Big]
\\& +u_+^2g\Big[F\big(\frac{\pi}{2},k\big)-F(\phi,k)\Big],
\end{split}
\end{align}
\begin{align}
\begin{split}
\boldsymbol{M_2} & = 3u_+\int_0^{u_2}du\frac{u}{\sqrt{-(u-u_1)(u-u_2)(u-u_3)(u-u_4)}}\\ &
=3u_+g\Big[\Big(u_1-u_4\Big)\Big(\Pi(\frac{\pi}{2},\alpha^2,k)-\Pi(\phi,\alpha^2,k)\Big)+u_4\Big(F(\frac{\pi}{2},k)-F(\phi,k)\Big)\Big],
\end{split}
\end{align}
and 
\begin{align}
\begin{split}
\boldsymbol{M_3} & = u_+^3\int_0^{u_2}du\frac{1}{(u-u_+)\sqrt{-(u-u_1)(u-u_2)(u-u_3)(u-u_4)}}\\ &
=u_+^3\frac{g}{(u_+-u_1)}\Big[\frac{u_1-u_4}{u_+-u_4}\Big(\Pi(\phi,\alpha_3^2,k)-\Pi(\frac{\pi}{2},\alpha_3^2,k)\Big)+\frac{u_+-u_1}{u_+-u_4}\Big(F(\phi,k)-F(\frac{\pi}{2},k)\Big)\Big]\,.
\end{split}
\end{align}
Therefore the integral $I_2$ can be written as,
\begin{align}
\begin{split}
& I_2=\int_0^{u_2}du \frac{u^3}{(u-u_+)\sqrt{-(u-u_1)(u-u_2)(u-u_3)(u-u_4)}} \\ &
=g\Big[\Big(u_4^2-2u_+u_4+u_+^2-\frac{u_+^3}{u_+-u_4}\Big)\Big(F(\frac{\pi}{2},k)-F(\phi,k)\Big)\\ &
+\Big(u_+(u_1-u_4)-\frac{u_+^3(u_1-u_4)}{(u_+-u_1)(u_+-u_4)}\Big)\Big(\Pi(\frac{\pi}{2},\alpha_3^2,k)-\Pi(\phi,\alpha_3^2,k)\Big)\\ &
+2u_4(u_1-u_4)\boldsymbol{V_1}+(u_1-u_4)^2\boldsymbol{V_2}\Big]\label{812}
\end{split}
\end{align}
where $\boldsymbol{V_1}$ and $\boldsymbol{V_2}$ is given by,
\begin{align}
\begin{split}
& \boldsymbol{V_1}=\Pi(\frac{\pi}{2},\alpha^2,k)-\Pi(\phi,\alpha^2,k)\,,\\ &
\boldsymbol{V_2}=\frac{1}{2(\alpha^2-1)(k^2-\alpha^2)}\Big[\alpha^2\Big(E(\frac{\pi}{2},k)-E(\phi,k)\Big)+(k^2-\alpha^2)\Big(F(\frac{\pi}{2},k)-F(\phi,k)\Big)\,,\\ &
(2\alpha^2\,k^2+2\alpha^2-\alpha^4-3k^2)\Big(\Pi(\frac{\pi}{2},\alpha^2,k)\Big)-\Pi(\phi,\alpha^2,k)\Big)\Big)-\frac{\alpha^4\,\sin\,\phi\sqrt{1-\sin^2\phi}\sqrt{1-k^2\,\sin^2\phi}}{1-\alpha^2 \,\sin^2\phi}\Big]
\end{split}
\end{align}
and
\begin{align}
    \begin{split}
   & g=\frac{2}{\sqrt{(u_4-u_2)(u_3-u_1)}},\,\, \alpha^2=\frac{u_1-u_2}{u_4-u_2}<0,\,\,
     \,\,
      \alpha_3^2=\alpha^2\frac{u_+-u_4}{u_+-u_1}\,,\\&
       k^2=\frac{(u_4-u_3)(u_2-u_1)}{(u_4-u_2)(u_3-u_1)},\,\,
       \phi=\arcsin\sqrt{-\Big(\frac{(u_4-u_2)u_1}{(u_4-u_1)u_4}\Big)}.
\end{split}
\end{align}

\bibliographystyle{utphysmodb}
\bibliography{ref}

\end{document}